\pdfoutput=1
\RequirePackage{rotating}
\documentclass[fleqn,usenatbib]{mnras}  

\usepackage[T1]{fontenc}
\usepackage{ae,aecompl}

\usepackage{mathptmx}

\usepackage{graphicx}	
\usepackage{amsmath}	
\usepackage{amssymb}	
\usepackage[british]{babel}
\usepackage[]{units}
\usepackage{multirow}
\usepackage{afterpage}
\usepackage{pdflscape}
\usepackage{rotating}
\usepackage{caption}
\usepackage{booktabs}
\usepackage[para,flushleft,referable]{threeparttablex}
\usepackage{xcolor}
\usepackage{microtype}
\hypersetup{pdfauthor={Errani et al.},
            pdftitle={Systematics in virial mass estimators for pressure-supported systems},
            pdfkeywords={Local Group, galaxies: kinematics and dynamics, galaxies: dwarf, dark matter},
            bookmarksnumbered=true}

\newcommand{\rmax}{r_\mathrm{max}}	%
\newcommand{\vmax}{v_\mathrm{max}}	%
\newcommand{\kms}{\mathrm{km\,s^{-1}}}		%
\newcommand{\Rh}{R_\mathrm{h}}	%
\newcommand{\declog}{\mathrm{log_{10}}}	%

\title[Systematics in virial mass estimators]{Systematics in virial mass estimators for pressure-supported systems}

\author[Errani et al.]{Rapha\"el Errani$^{1}$\thanks{E-mail: raer@roe.ac.uk}, Jorge Pe\~narrubia$^{1}$ \& Matthew G. Walker$^{2}$
\\
{$^1$ Institute for Astronomy, University of Edinburgh, Royal Observatory, Blackford Hill, Edinburgh EH9 3HJ, UK}\\
{$^2$ McWilliams Center for Cosmology, Department of Physics, 5000 Forbes Ave., Carnegie Mellon University, Pittsburgh, PA 15213, USA}
}

\date{Accepted 2018 September 11. Received 2018 August 17; in original from 2018 May 1}

\pubyear{2018}
\volume{}

\begin{document}

\label{firstpage}
\pagerange{\pageref{firstpage}--\pageref{lastpage}} \pubyear{2018}
\maketitle

\begin{abstract}
Mass estimators are a key tool to {infer the dark matter content in pressure-supported systems like dwarf spheroidal galaxies (dSphs)}. We construct an estimator for enclosed masses based on the virial theorem which is insensitive to anisotropy in the velocity dispersion and tailored to yield masses with minimum uncertainty introduced by our ignorance on (i) the shape of the inner halo profile, and (ii) how deeply the stellar component is embedded within the halo: \mbox{$M(<1.8\,R_\mathrm{h}) \approx 3.5 \times 1.8\,R_\mathrm{h} \langle \sigma_\mathrm{los}^2 \rangle G^{-1}$}, where by $\Rh$ we denote the projected half-light radius and by $\langle \sigma_\mathrm{los}^2 \rangle$ the luminosity-averaged squared line-of-sight velocity dispersion. 
Tests against controlled simulations show {that this estimator provides unbiased enclosed masses} with an accuracy of $\sim 10$ per cent. {This confirms the robustness of similar previously proposed mass estimators.}
Application to published kinematic data of Milky Way dSphs reveals a tight correlation between enclosed mass and luminosity. Using $N$-body models we show that tidal stripping has little effect on this relation.  Comparison against cuspy and cored dark matter haloes extracted from controlled re-simulations of the Aquarius A2 merger tree show{s} that the high mass densities of ultrafaint galaxies are not compatible with large dark matter cores, and that the (total) halo masses of the {classical} Milky Way dSphs span a remarkably narrow range (\mbox{$8 \lesssim \declog\,(M/\mathrm{M_\odot}) \lesssim 10$}) at present, showing no clear trend with either galaxy size or luminosity. 
\end{abstract}



\section{Introduction}

Understanding the distribution of dark matter (DM) on galactic scales is at the root of constraining DM particle properties and processes of {galaxy formation like baryonic feedback}. Dwarf spheroidal galaxies (dSphs), lying at the faintest end of the galaxy luminosity function and reaching mass-to-light ratios of up to $\sim10^3$ \citep[see e.g.][]{Walker2007,McConnachie2012} are promising candidates to study the DM distribution on kpc scales. But obtaining such constraints through kinematic data is a challenging task: the galaxies are observed in projection, and precise velocity measurements are feasible only along the line of sight - though recently attempts have been made to constrain the 3D motion of stars in the Sculptor dwarf galaxy \citep{Massari2018}. 

{
Having access only to velocities along the line of sight, the enclosed mass $M(<r_0)$ within a specific spherical radius $r_0$ may be obtained from the projected virial theorem for spherical systems \citep[see e.g.][]{MerrifieldKent1990,AgnelloEvans2012}, relating the projected components of the pressure and potential energy tensors.
For the case of self-gravitating systems and those where mass traces light with De Vaucouleurs luminosity profile, \citet{Illingworth1976} describes how to infer the total mass ($r_0 \rightarrow \infty$) from the luminosity-averaged squared line-of-sight velocity dispersion $\langle\sigma^2_\mathrm{los}\rangle$. \citet{Merritt1987} shows how lower limits on the enclosed mass $M(<r_0)$ can be obtained from the virial theorem for systems where mass and luminosity profile have different functional forms.
Mass estimates from the virial theorem can be obtained also for dSphs consisting of a stellar tracer embedded in a DM halo (see section \ref{sec:massEstimates}). Such estimates are based entirely on projected and averaged dispersion measurements, though including velocity components tangential to the line of sight is straightforward (see section \ref{sec:conclusionsSummary}). 
The method requires however to make assumptions on the DM halo shape and on how deeply the stellar component is embedded within the halo. 

Alternatively, masses enclosed within a spherical radius $r_0$ can be obtained from the radial velocity dispersion $\sigma_\mathrm{r}(r)$ at a radius $r=r_0$ by means of the Jeans equations \citep[see e.g.][]{BT87}.
Both the radial velocity dispersion $\sigma_\mathrm{r}(r)$ at a spherical radius $r$ and the observed line-of-sight velocity dispersion $\sigma_\mathrm{los}(R)$ measured at a projected radius $R$ depend on the DM halo shape as well as the function $\beta(r) \equiv 1 - {\sigma^2_{\bot}(r)}/{\sigma^2_{r}(r)}$ which parametrizes the anisotropy of the velocity dispersion as a function of radius $r$, $\sigma^2_{\bot}(r)$ being the tangential component of the velocity dispersion \citep[see e.g.][]{BinneyMamon1982}.
Current galaxy formation models do not yield clear predictions for the behaviour of $\beta(r)$. In principle, $\beta(r)$ does not need to be a monotonic function of $r$, and it might vary between different stellar populations within the same dwarf. Our ignorance of $\beta(r)$ gives rise to the infamous mass - anisotropy degeneracy.}

{
Various mass estimators have been proposed which rely on the assumption that the enclosed mass is directly proportional to local or global averages of the observed line-of-sight velocity dispersion. Mass estimators of this type have been tested on cosmological simulations \citep{Campbell2016, GonzalesSamaniego2017} and proven to give accurate masses within a factor of unity.}
E.g., for the case of isotropic (i.e. $\beta(r)=0$) King models embedded in \citet{nfw1997} haloes, \citet{PenarrubiaMcConnachieNavarro2008} show that measuring the mass $M(<r_0)$ enclosed within the projected King core radius $r_0 = R_\mathrm{c}$ minimizes the uncertainty introduced by how deeply the stellar component is segregated within the DM halo. Their estimator reads $  M( < R_\mathrm{c} ) \approx   {1.44\,R_\mathrm{c}~ \sigma_\mathrm{los}^2(0) }{G^{-1}}$, where $\sigma_\mathrm{los}^2(0)$ denotes the central line-of-sight velocity dispersion.
\citet{Walker2009} model the stellar tracer profiles as Plummer spheres and assume flat velocity dispersion profiles to estimate the enclosed mass as $  M( < R_\mathrm{h} ) \approx   {2.5\,R_\mathrm{h}~ \langle \sigma_\mathrm{los}^2 \rangle }{G^{-1}}$, showing numerically that the uncertainty introduced by a constant $\beta(r)$ is relatively small when measuring the enclosed mass at the projected half-light radius $\Rh$. \citet{Wolf2010} reach a similar conclusion by analytical means, showing that the deprojected half-light radius approximately coincides with the radius where the anisotropy parameter $\beta(r)$ (modelled using a three-parameter function) has the smallest effect on the inferred enclosed mass. Their estimator reads $ M( < r_\mathrm{h} ) \approx   {3.0\,r_\mathrm{h}~ \langle \sigma_\mathrm{los}^2 \rangle  }{G^{-1}}$, where $r_\mathrm{h} \approx 4/3 ~ R_\mathrm{h}$ is the deprojected half-light radius.
For the family of Michie-King distribution functions, \citet{Amorisco2012} find that $  M( < 1.7\,\Rh ) \approx  {5.8\,\Rh~ \sigma_\mathrm{los}^2(R_\mathrm{h})  }{G^{-1}} $, where $\sigma_\mathrm{los}^2(R_\mathrm{h})$ is measured {at the projected half-light radius $\Rh$}. By purely numerical means, \citet{Campbell2016} find $  M( < 1.8\,\Rh ) \approx  {6.0\,R_\mathrm{h}~ \langle \sigma_\mathrm{los}^2(<1.04\,R_\mathrm{h}) \rangle }{G^{-1}} $ from a dispersion-supported sample of galaxies of the \textsc{apostle} simulation project \citep{SawalaFrenk2016}, averaging $\sigma^2_\mathrm{los}$ within $1.04\, R_\mathrm{h}$, {denoting by $\Rh$ the projected half-light radius.}

The mass - anisotropy degeneracy may be broken if kinematic measurements of several chemo-dynamically distinguishable stellar populations within the same dSph can be obtained as discussed e.g. by
\citet{Saglia2000}, \citet{Battaglia2008} and \citet{Amorisco2012}, whereas \citet{Diakogiannis2014a} suggest to measure $\sigma_\mathrm{los}(R)$ at different projected radii $R$ and to model $\beta(r)$ using splines.
With access to proper motions tangentially to the line of sight for a subsample of tracer stars, stronger constraints on $\beta(r)$ can be obtained \citep[see e.g.][]{StrigariBullockKaplinghat2007,Watkins2013}.
\citet{LokasMamon2003} show that a joint analysis of the velocity dispersion profile and higher order moments like the kurtosis may be used to break the mass - anisotropy degeneracy under specific assumptions for the form of $\beta(r)$. \citet{BreddelsHelmi2013} model dSphs by Schwarzschildt orbit superposition to fit both the observed light distribution as well as the second- and forth-order velocity moments, and apply the method to a catalogue of 2000 stars of the Sculptor dSph. They show that with 2000 tracers, the central slope {$\gamma=-\mathrm{\mathrm{d}\,ln\,}\varrho(r)/{\mathrm{d}\,\ln\,} r\,(r\rightarrow0)$ of the DM density profile $\varrho(r)$} cannot be constrained to tell cuspy from cored profiles, and $\beta(r)$ cannot be constrained for radii smaller than \unit[0.1]{kpc} - these authors conclude that to obtain unbiased estimates {of the central slope $\gamma$}, a larger number of tracer stars is necessary. \citet{RichardsonFairbairn2014} use virial shape parameters derived from the fourth-order virial equations to constrain DM density profiles compatible with the observed kinematics of embedded stellar tracers. Applied to the Sculptor dSph, the authors find that a cuspy DM halo is favoured if the stellar distribution is modelled as a Plummer profile. If the outer slope of the stellar profile however is allowed to vary, the available kinematic data do not allow to constrain the central DM profile slope.
Using $N$-body mock data, \citet{Read2017} show that with $10^4$ tracer stars, no reliable $\beta(r)$ profile can be recovered from projected velocities and Jeans analysis alone. However, making use also of virial shape parameters or velocity measurements tangentially to the line of sight, using $10^3$ tracer stars, both $\beta(r)$ and the underlying mass profile can be constrained within a few multiples of the half-light radius. A different approach is discussed by \citet{Wang2015}, who fit analytical distribution functions to observables extracted from stellar tracers embedded in DM haloes of the Aquarius \citep{Springel2008} simulations. These authors find that the derived halo masses are biased (by up to 40 per cent) even when using not only line of sight but also tangential velocity measurements.

Mass estimators played a crucial role in identifying possible tensions between $N$-body simulations of Milky Way-like DM haloes and the observed population of satellite galaxies of the Milky Way: \citet{BoylanKolchin2011} made use of the \citet{Wolf2010} estimator to identify $N$-body subhaloes compatible with the mean densities of Milky Way dwarfs, finding that none of the bright Milky Way dwarfs are dense enough to be embedded in haloes similar to the most massive simulated $N$-body subhaloes of the Aquarius simulations (the \emph{too big to fail problem}). Based on estimates of enclosed masses of two distinct stellar populations embedded in the same DM halo, \citet{Walker2011} introduced a method to infer {the slope \mbox{$\Gamma \approx \mathrm{\Delta\,ln\,}M(<r)/{\Delta\,\ln\,} r\,(r\rightarrow\bar R_{\mathrm{h}})$} of the underlying DM halo at the average half-light radius $\bar R_{\mathrm{h}}$}, finding that both the Fornax and the Sculptor dSph must be embedded in DM haloes with constant-density cores, in conflict with the centrally divergent density cusps of haloes in DM-only simulations (the \emph{cusp/core problem}). \citet{Laporte2013} subsequently showed that also for triaxial DM haloes, the \citet{Walker2011} method gives reliable lower limits on the slope~$\Gamma$.
\citet{Sawala2015} use halo masses derived by \citet{PenarrubiaMcConnachieNavarro2008} for Milky Way dSphs using the \citet{Walker2009} estimator to motivate their corrections to stellar mass - halo mass relations obtained from 
abundance matching based on DM-only simulations.
More recently, \citet{Fattahi2018} make use of the \citet{Wolf2010} estimator to study the tidal evolution DM haloes of the \textsc{apostle} simulations compatible with the observed kinematics of Milky Way dSphs.

In this contribution, we study a virial mass estimator which is minimally affected by our ignorance on two key parameters: (i)~the shape of the inner DM halo profile, and (ii) the spatial segregation of the stellar tracers within the DM halo. The use of the projected spherical virial theorem has several advantages over Jeans analysis for the construction of mass estimators: 
The virial theorem does not contain any functional dependence on $\beta(r)$ and thereby avoids the mass-anisotropy degeneracy. Also systematic biases of inferred enclosed masses and slopes $\Gamma$ follow directly from the assumptions made on the DM and stellar density profiles and are independent of sample size. Furthermore, the average squared dispersion $\langle \sigma_\mathrm{los}^2 \rangle$ is a sum over all stars and does not require any binning of data. This makes mass estimators based on the virial theorem applicable also to systems with only few stellar tracers - carefully factoring in the uncertainties due to sample size on the measured squared velocity dispersion as pointed out by \citet{Laporte2018}.

This paper is structured as follows: in section \ref{sec:massEstimates} we introduce the spherical virial theorem and discuss the systematics of mass estimators for dwarf galaxies consisting of a stellar tracer population embedded inside a DM halo. In section \ref{sec:unbiased} we construct a mass estimator which is tailored to minimize the uncertainty introduced by our ignorance on the central slope of the underlying DM profile as well as on how deeply embedded the stellar component is within the DM halo. We do test this mass estimator on a catalogue of $N$-body mocks. Section \ref{sec:totalmass} discusses how to obtain the total halo masses of dwarf galaxies, breaking the degeneracy between structural parameters constrained by use of the virial theorem with the help of controlled cosmological simulations. 
In section \ref{sec:MWdwarfs}, we estimate the masses of Milky Way dwarf galaxies using the methods introduced in the previous sections, and discuss implications of the derived stellar mass - halo mass relation for satellite galaxies. Section \ref{sec:conclusionsSummary} summarizes methods and main results.

\section{Mass estimates from the virial theorem}
\label{sec:massEstimates}
With the aim of estimating masses of dwarf galaxies from their observed velocity dispersions while avoiding the infamous mass - anisotropy degeneracy,
in the spirit of \citet{AgnelloEvans2012}, we construct a mass estimator based on the projected virial theorem for spherical systems,
\begin{equation}
\label{eq:virial}
 2K_\mathrm{los} + W_\mathrm{los} = 0 ~.
\end{equation}
We distinguish between the mass profile $M(<r)$ which sources the potential, and a mass-less tracer component of density $\nu_\star(r)$ embedded in this potential.
The pressure term is given by the projected velocity dispersion of the tracer and can be measured directly:
\begin{equation}
\label{eq:pressureTerm}
2K_\mathrm{los} = 2\pi  \int_0^\infty \Sigma_\star(R) \, \sigma^2_\mathrm{los}(R) \, R\, \mathrm{d}R \equiv \langle \sigma_\mathrm{los}^2 \rangle ~,
\end{equation}
where by $\Sigma_\star(R)$ and $\sigma^2_\mathrm{los}(R)$ we denote the surface density and line-of-sight velocity dispersion of the tracer {at a projected (2D) radius $R$, respectively. The surface density is normalized so that \mbox{$2\pi\int_0^\infty R\Sigma_\star(R) \mathrm{d}R =1$}. }
The potential energy term
\begin{equation}
\label{eq:potTerm}
 W_\mathrm{los} = - \frac{4 \pi G}{3} \int_0^\infty r \nu_\star(r) M(<r) \mathrm{d}r 
\end{equation}
can be calculated for a given normalized tracer density profile $\nu_\star(r)$ and mass distribution $M(<r)$, where {$r$ is the spherical (3D) radius and the tracer density is normalized so that \mbox{$4\pi\int_0^\infty r^2 \nu_\star(r) \mathrm{d}r =1$}. We will keep this notation for spherical ($r$) and projected ($R$) radii through the paper.}
Equations \ref{eq:virial} - \ref{eq:potTerm} allow $\langle \sigma_\mathrm{los}^2 \rangle = - W_\mathrm{los}$ to be calculated from the tracer and mass profiles alone. In contrast to the Jeans equations, these virial equations and in specific the integral $\langle \sigma_\mathrm{los}^2 \rangle$ are independent of anisotropies in the tracer velocity dispersion and thereby avoid the mass - anisotropy degeneracy.

The mass within a radius $r_0$ can be computed from the observed $\langle \sigma_\mathrm{los}^2 \rangle$ as $ M(<r_0) = {\mu ~r_0 ~\langle \sigma_\mathrm{los}^2 \rangle }/{G} $
where 
\begin{equation}
\label{eq:mu-general-form}
 \mu(r_0) = - \frac{G~M(<r_0)}{r_0~W_\mathrm{los} }
\end{equation}
is a dimensionless function of the radius $r_0$ and the parameters describing the tracer density and mass distribution.

\subsection{Self-gravitating systems}
As a first example, let us consider the case of a spherical system where the tracer density equals the mass density.
For \citet{dehnen1993} profiles of total mass $M_\mathrm{tot}$, scale radius $a$, scale density $\varrho_s = (3-\gamma)M_\mathrm{tot}/4\pi a^3$, outer slope $\beta\equiv-\mathrm{d}\ln \varrho/\mathrm{d}\ln r~(r\rightarrow \infty)=4$, inner slope $\gamma\equiv-\mathrm{d}\ln \varrho/\mathrm{d}\ln r~(r\rightarrow 0)$ within the range $0 \leq \gamma \leq 1$, written in terms of the general $\{\alpha,\beta,\gamma\}$ profile with $\alpha=1$,
\begin{equation}
\label{eq:betagammaprofile}
 \varrho(r) = \varrho_s \left( \frac{r}{a} \right)^{-\gamma} \left[ 1 + \left( \frac{r}{a} \right)^\alpha \right]^{(\gamma-\beta)/\alpha}~,
\end{equation}
the potential energy term becomes
\begin{equation}
 W_\mathrm{los} = - \frac{GM_\mathrm{tot}}{a} \frac{1}{6(5-2\gamma)}
\end{equation}
and
\begin{equation}
 \mu_\gamma(r_0/a) = \left (  1+ \frac{a}{r_0} \right)^{\gamma-3} \left( \frac{a}{r_0} \right)   6(5-2\gamma)   
\end{equation}
is a function of the radius $r_0$ and the scale radius $a$ in the combination $r_0/a$ alone. For a fixed $r_0$, $\mu_\gamma$ reduces to a number, e.g. for the deprojected half-light radius $r_{\mathrm{h}}= a /(2^{1/(3-\gamma)} -1 ) $, $\mu_1(r_{\mathrm{h}}) = 9 (\sqrt{2}-1) \approx 3.7$, and $\mu_0(r_{\mathrm{h}}) = 15(2^{1/3}- 1) \approx 3.9$.

{
\citet{Illingworth1976} introduced an estimator based on the same approach for the total mass of self-gravitating elliptical galaxies which follow a De Vaucouleurs surface brightness profile, i.e. $\log\, \Sigma_\star(R)/\Sigma_\star(0) \propto R^{1/4}$. The estimator reads
$M_\mathrm{tot} \approx 8.6\,\Rh\langle \sigma_\mathrm{los}^2 \rangle / G$, denoting by $\Rh$ the projected half-light radius.
In comparison, for our example of \citet{dehnen1993} density profiles, the projected half-light radius $\Rh$ equals $\approx 2.9a$ ($1.8a$) for $\gamma=0$ ($\gamma=1$), and we find $M_\mathrm{tot} \approx 10.3\,\Rh\langle \sigma_\mathrm{los}^2 \rangle / G$ for $\gamma=0$ and $M_\mathrm{tot} \approx 9.9\,\Rh\langle \sigma_\mathrm{los}^2 \rangle / G$ for $\gamma=1$.
Using Schwarzschildt orbit modelling under the assumption that mass follows light, \citet{Cappellari2006} fit kinematics of 25 elliptical (E) and lenticular (S0) galaxies to obtain the total mass estimate $M_\mathrm{tot} \approx 5\, \Rh \langle \sigma_\mathrm{los}^2 (<\Rh)\rangle / G$, where $\langle \sigma_\mathrm{los}^2 (<\Rh)\rangle$ is averaged within the projected half-light radius $\Rh$. Note that this estimator is not independent of the anisotropy profile $\beta(r)$ as only the virial average $\langle \sigma_\mathrm{los}^2\rangle$  over the entire system removes the dependence on $\beta(r)$. \citet{Agnello2014} argue that this global average can be challenging to compute for faint systems, and explore for systems with De Vaucouleurs luminosity profile and Ospikov-Meritt anisotropy profile $\beta(r) = r^2/(r^2+r_a^2)$ how the coefficient $GM(<R)/R\langle \sigma_\mathrm{los}^2(<R)\rangle$ changes as a function of the radius $R$ over which the velocity dispersion is averaged. For systems where mass traces light, they propose a mass estimator similar to \citet{Illingworth1976}.

}

\subsection{Stellar tracer embedded in DM potential}
For the case of dwarf galaxies with a stellar population embedded inside a DM halo, the mass profile which sources the potential is not accessible to direct observation.
Let us assume that the projected half-light radius $\Rh$ and the luminosity-averaged squared velocity dispersion $\langle \sigma_\mathrm{los}^2 \rangle$ are known quantities. 
This motivates a virial mass estimator for the enclosed mass $M(<r_0)$ with $r_0=\lambda \Rh$ of the form \citep{Amorisco2012}:
\begin{equation}
\label{eq:massEst}
M_\mathrm{est}( < \lambda R_\mathrm{h} ) =   \frac{\mu~\lambda R_\mathrm{h}~ \langle \sigma_\mathrm{los}^2 \rangle }{G}  ~,
\end{equation}
where $\lambda$ and $\mu$ are dimensionless factors. 

Using equation~\ref{eq:massEst}, we now aim to determine the mass enclosed within the spherical radius $r_0=\lambda \Rh$. 
We treat stars as mass-less tracers of the underlying DM potential, consequently \mbox{$M(<r)$} is determined by the DM profile alone. 
In the following, we assume that the stellar tracer distribution follows a spherical Plummer profile, $\{\alpha_\star,\beta_\star,\gamma_\star\}=\{2,5,0\}$.
To develop some intuition for the range of values which $\mu$ can take, let's consider a (scale-free) power-law DM profile with $\varrho \propto r^{-\gamma}$, i.e. the asymptotic case $R_\mathrm{h}/\rmax \rightarrow 0$ of a stellar population deeply embedded inside the DM host halo. Then $\mu$ is a function of $\lambda$ and $\gamma$ alone, 
\begin{equation} 
\label{eq:asymptoticMu}
\mu_\gamma(\lambda)  \stackrel{R_\mathrm{h} \rightarrow 0}{=} ~~  \frac{2 \lambda^{2-\gamma}}{\mathrm{B}\left({\gamma}/{2},\,(5-\gamma)/{2}\right)} 
\end{equation}
where by $\mathrm{B}$ we denote the Euler beta function. For $\gamma=1$ this reduces to $\mu_1 = 3 \lambda / 2$, whereas for $\gamma \rightarrow 0$, $\mu_\gamma \rightarrow 0$ as $\langle \sigma^2_\mathrm{los}\rangle = - W_\mathrm{los}$ diverges. { For scale-free cuspy DM haloes with $1<\gamma<3$, similar results are shown in \citet[][figures 2 \& 3]{Agnello2014} computed from the Jeans equations for stellar tracers with De Vaucouleurs luminosity profile and Ospikov-Merritt anisotropy profile.
}

\begin{figure}
  \centering
  \includegraphics[width=8.5cm]{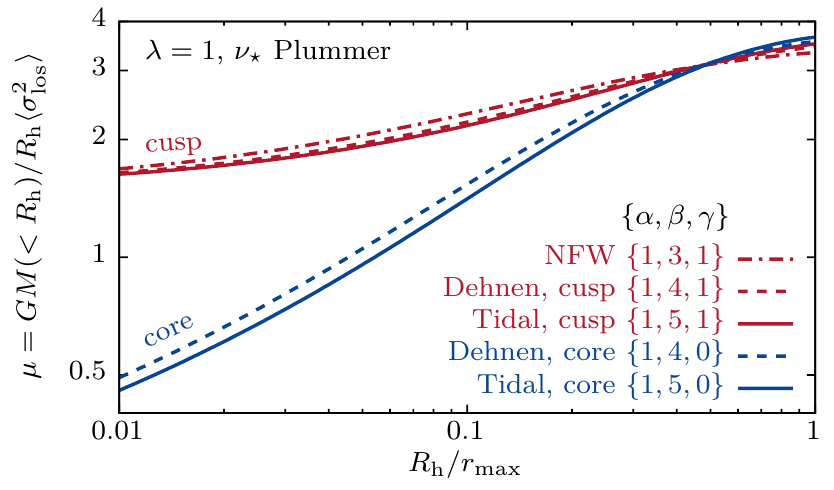}
  \caption{Dimensionless factor $\mu$ for $\lambda=1$ (see equation~\ref{eq:massEst}) as a function of segregation $R_\mathrm{h}/\rmax$ for stellar Plummer profiles with projected half-light radius $R_\mathrm{h}$ embedded in cored ($\gamma=0$) and cuspy ($\gamma=1$) DM haloes for tidally stripped systems, $\{\alpha,\beta,\gamma\}=\{1,5,\gamma\}$ (see equation~\ref{eq:betagammaprofile}). For comparison, the functional dependence of $\mu$ on segregation is also shown for cuspy and cored Dehnen ($\beta=4$) and NFW ($\beta=3$) profiles.  }
  \label{fig:muLambda1}
\end{figure}
For non scale-free DM haloes, e.g. the density profile of equation~\ref{eq:betagammaprofile} with scale radius $a$ and two different asymptotic behaviours for $r \gg a$ and $r \ll a$, $\mu$ is not only a function of $\lambda$ but also of the segregation parameter $\Rh/\rmax$ describing how deeply embedded the stars are within the DM halo. There, by $\rmax$ we denote the radius of maximum circular velocity $\vmax = \left(GM\left(<\rmax\right)/\rmax\right)^{1/2}$.
Motivated by the result of controlled simulations that DM haloes undergoing tidal stripping have { steeper outer slopes $\beta\equiv-\mathrm{d}\ln \varrho/\mathrm{d}\ln r~(r\rightarrow \infty)$} than field haloes, i.e. $\beta \approx 5$ \citep[see][]{Penarrubia2009, Penarrubia2010} in contrast to the outer slope of $\beta=3$ of NFW profiles \citep{nfw1997}, Fig.~\ref{fig:muLambda1} shows $\mu$ for the choice of $\lambda = 1$ as a function of the segregation parameter $\Rh/\rmax$ for stellar Plummer spheres embedded in cored ($\gamma=0$) and cuspy ($\gamma=1$) DM haloes with density profile as in equation~\ref{eq:betagammaprofile}, using $\beta=5$ and $\varrho_s = M_\mathrm{tot} {(3-\gamma)(4-\gamma)}/{4 \pi a^3}$. We choose $\alpha=1$ motivated by the value found for equilibrium haloes in cosmological $N$-body simulations \citep{nfw1997}. For comparison, the functional behaviour of $\mu$ is also shown for Dehnen, i.e. $\{\alpha,\beta,\gamma\}=\{1,4,\gamma\}$, and NFW $\{1,3,1\}$ profiles. In the limit $R_\mathrm{h}/\rmax \rightarrow 0$, we recover the asymptotic behaviour of equation~\ref{eq:asymptoticMu} for $\lambda = 1$: $\mu \rightarrow 3/2$ for haloes with $\gamma=1$, and $\mu \rightarrow 0$ for cored ones. For the segregations shown, $\mu$ is less sensitive to the steepness of the outer slope ($\beta=3,4,5$) of the DM halo than it is to the inner one ($\gamma=0,1$).

Our ignorance of $\Rh/\rmax$ limits us to the use of a constant value of $\mu$ when estimating masses using equation~\ref{eq:massEst}. 
This leads to a systematic uncertainty on the estimated masses, and this uncertainty is more pronounced for cored haloes where $\mu_{\gamma=0}\rightarrow0$ for the case of deeply embedded stellar populations. In section \ref{sec:unbiased} we will discuss choices of $\mu,\lambda$ which minimize this uncertainty.

\subsection{Slope estimates from two stellar populations}

\begin{figure}
  \centering
  \includegraphics[width=8.5cm]{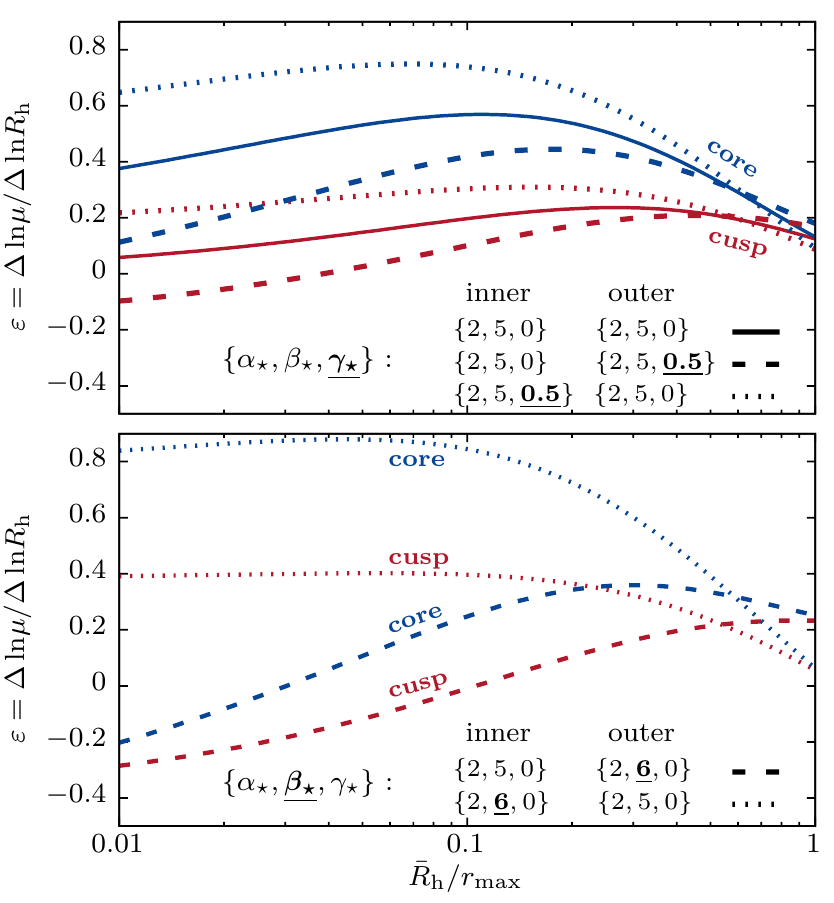}
  \caption{Logarithmic slope $\varepsilon = \Delta \ln\, \mu / \Delta \ln\, R_\mathrm{h}$ computed for two stellar tracer populations with half-light radii $R_{\mathrm{h},\mathrm{inner}}$ and $R_{\mathrm{h},\mathrm{outer}} = 2 R_{\mathrm{h},\mathrm{inner}}$, embedded in cuspy ($\gamma=1$) or cored ($\gamma=0$) DM haloes with $\{\alpha,\beta,\gamma\}=\{1,5,\gamma\}$ density profile (equation~\ref{eq:betagammaprofile}). The slope is shown as a function of the mean half-light radius $\bar R_\mathrm{h} = (R_{\mathrm{h},\mathrm{inner}} + R_{\mathrm{h},\mathrm{outer}})/2$. \emph{Top panel}: solid lines show $\varepsilon$ for systems where both the inner and the outer stellar tracer populations have Plummer $\{\alpha_\star,\beta_\star,\gamma_\star\}=\{2,5,0\}$ density profiles. The dashed curves correspond to systems where one of the tracer populations has a shallow central density cusp of slope $\gamma_\star$. \emph{Bottom panel:} the outer slope $\beta_\star$ of one of the tracers is chosen to be steeper than the Plummer case.}
  \label{fig:alphaLambda1}
\end{figure}

{
If we do have at hand two chemo-dynamically distinguishable stellar populations with projected half-light radii $R_\mathrm{h,inner}, R_\mathrm{h,outer}$ (where $R_\mathrm{h,inner} < R_\mathrm{h,outer}$) embedded in the same DM halo, the combined velocity dispersion measurements from both populations can be used to further constrain structural properties of the underlying DM halo. 
\citet{AgnelloEvans2012} use the virial theorem to obtain a lower limit on the core size of the DM halo of the Sculptor dSph, modelling the stellar populations as Plummer spheres and the DM halo as a cored NFW profile \citep[][equation 18]{AgnelloEvans2012}.

\citet{Walker2011} introduce an estimator for the central slope $\Gamma$ of the underlying DM halo which does not rely on assumptions about the DM profile shape:
\begin{equation}
 \Gamma = \frac{\mathrm{d} \mathrm{ln}M(<r)}{\mathrm{d}\mathrm{ln}r} \approx \underbrace{  \vphantom{1~+~\frac{2\,\mathrm{ln}\, \sigma_2/\sigma_1}{\mathrm{ln}\,R_{\mathrm{h},2}/R_{\mathrm{h},1}    } }   \frac{\mathrm{\Delta} \mathrm{ln}\,\mu}{\mathrm{\Delta}\mathrm{ln}\,R_\mathrm{h}} }_{\varepsilon} ~+~ \underbrace{1~+~\frac{2\,\mathrm{ln}\, \sigma_\mathrm{outer}/\sigma_\mathrm{inner}}{\mathrm{ln}\,R_{\mathrm{h},\mathrm{outer}}/R_{\mathrm{h},\mathrm{inner}}    } }_{\mbox{WP11}} ~,
\end{equation} 
where WP11 refers to the estimator introduced by \citet{Walker2011}, and writing for brevity $\sigma \equiv \langle \sigma_\mathrm{los}^2 \rangle^{1/2}$. The virial theorem allows us to study the systematics of this estimator for different DM and stellar density profiles. For this aim, we embed two stellar tracer populations with density $\nu_\star(r)$
and projected half-light radii $R_{\mathrm{h,inner}}$ and $R_{\mathrm{h,outer}} = 2\, R_{\mathrm{h,inner}}$ in cuspy ($\gamma=1$) or cored ($\gamma=0$) tidally stripped DM haloes (equation~\ref{eq:betagammaprofile} with $\alpha=1,\beta=5$). The top panel of Fig.~\ref{fig:alphaLambda1} shows the resulting $\varepsilon= \Delta \ln \mu / \Delta \ln R_\mathrm{h}$ as a function of segregation.}
\begin{itemize}
 \item If $\nu_{\star,\mathrm{inner}}$ and $\nu_{\star,\mathrm{outer}}$ are Plummer profiles, $\{\alpha_\star,\beta_\star,\gamma_\star\} = \{2,5,0\}$, (solid lines), then $\varepsilon > 0$ for the range of segregations $R_\mathrm{h}/\rmax$ shown, i.e. in this case WP11 underestimates the slope $\Gamma$ of the underlying DM profile with $\varepsilon \lesssim 0.2,0.6$ for the cuspy and cored DM halo, respectively. 
 \item WP11 also underestimates $\Gamma$ if the inner population has a shallow central density cusp, $\{\alpha_\star,\beta_\star,\gamma_\star\} = \{2,5,0.5\}$, and the outer population follows a Plummer density profile (dotted lines).
 \item If instead the outer population has a shallow central density cusp and the inner population is a Plummer profile, then $\varepsilon < 0$ for cuspy haloes and segregations $R_\mathrm{h}/\rmax \lesssim 0.1$, i.e. in this configuration, WP11 overestimates $\Gamma$ (dashed lines).
\end{itemize}
In the bottom panel of Fig.~\ref{fig:alphaLambda1}, the outer slope of one of the stellar tracers is chosen to be steeper, $\{\alpha_\star,\beta_\star,\gamma_\star\} = \{2,6,0\}$, than in the Plummer case:
\begin{itemize}  
 \item If the outer population has a steeper outer slope $\beta_\star$ whereas the inner population is a Plummer profile, then $\varepsilon<0$ for deeply embedded tracers (dashed lines).
 \item On the other hand, if the inner population has a steeper outer slope and the outer population is a Plummer profile, $\varepsilon$ takes values of up to order of unity for deeply embedded tracers in cored DM haloes (dotted lines).
\end{itemize}
These results show that the shapes of the stellar tracer profiles need to be taken into account when estimating $\Gamma$, as omitting to do so may lead to underestimation of $\Gamma$ by values of up to order of unity.

\section{Unbiased minimum variance mass estimates}
\label{sec:unbiased}
In the previous section, we have shown how our ignorance on the segregation parameter $\Rh/\rmax$ leads to an uncertainty on the estimated masses which depends on the shape of the underlying DM profile.
We now aim to determine values $\bar \lambda, \bar \mu$ so that the mass estimator of equation~\ref{eq:massEst} is unbiased regarding (i) the DM profile shape (ii) how deeply embedded the stellar population is within the DM halo. By unbiased we intend not to make prior assumptions about the specific value of the segregation parameter $\Rh/\rmax$  and the inner\footnote{As shown in Fig.~\ref{fig:muLambda1}, comparing the cuspy Dehnen profile with outer slope $\beta=4$ to the NFW profile ($\beta=3$), for deeply embedded stellar systems, the outer slope of the DM halo does not affect the value of $\mu$ by much} slope $\gamma$ of the DM potential. By minimum variance we intend to choose $\bar \lambda, \bar \mu$ so that the relative error on the estimated masses due to our ignorance on $\Rh/\rmax$ and $\gamma$ is minimized.

\subsection{Minimizing the variance as a function of $\lambda$}
\begin{figure}
  \centering
  \includegraphics[width=8.5cm]{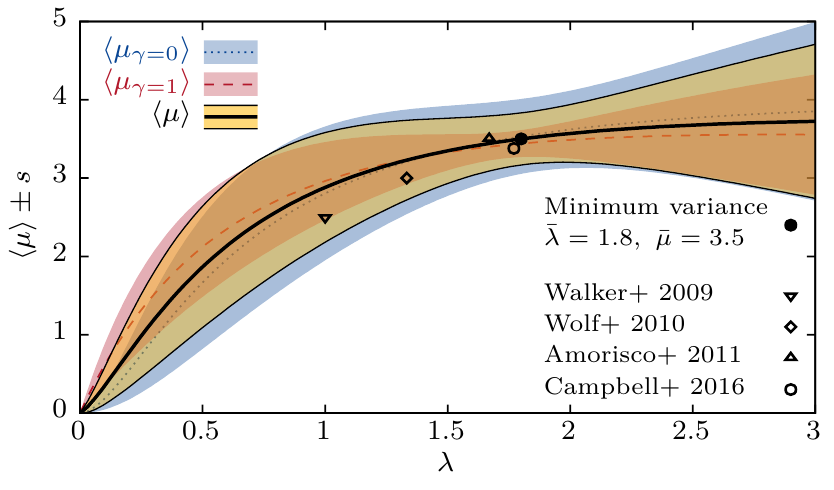}
  \caption{$\langle \mu_\gamma(\lambda) \rangle \pm s$ marginalized over segregation \mbox{$0\leq R_\mathrm{h}/\rmax\leq 1$} as a function of $\lambda$ for stellar Plummer spheres with projected half-light radius $R_\mathrm{h}$ embedded in cored ($\gamma=0$, blue dotted curve and blue shaded area) and cuspy ($\gamma=1$, red dashed curve and red shaded area) $\{\alpha,\beta,\gamma\}=\{1,5,\gamma\}$ DM profiles (equation~\ref{eq:betagammaprofile}). The black curve and yellow shaded area show $\langle\mu\rangle \pm s$ marginalized not only over segregation, but also over $0<\gamma<1$. The values $\bar \lambda,\bar \mu$ for our minimum variance estimator are indicated by a filled circle, whereas literature values for $\lambda, \mu$ are plotted using open symbols.}
  \label{fig:lambdaBias}
\end{figure}

Under the assumption of stellar Plummer spheres { which are deeply embedded inside tidally stripped DM haloes} ($\alpha=1,\beta=5$ in equation~\ref{eq:betagammaprofile}), for a given value of $\lambda$, we  marginalize $\mu$ (equation~\ref{eq:massEst}) over segregations in the range \mbox{$0\leq R_\mathrm{h}/\rmax\leq 1$} and central DM slopes in the range \mbox{$0 \leq \gamma \leq 1$}. We assuming flat distributions when marginalizing over $\Rh$ and $\rmax$:
\begin{equation}
\label{eq:int}
\langle \mu(\lambda) \rangle = \int_0^{1} \mathrm{d}(R_\mathrm{h}/\rmax)  \int_0^1 \mathrm{d}\gamma ~ \mu_\gamma(\lambda,R_\mathrm{h}/\rmax) ~~.
\end{equation}
{ The steepest central slope considered corresponds to the central slope of the NFW profile, and the range of slopes is consistent with the slopes measured in the hydrodynamical simulations of the MaGICC project \citep{DiCintio2014}.}
The upper limit of the integration range in $R_\mathrm{h}/\rmax$ is motivated by the findings reported in section \ref{sec:dwarf} from a sample of Milky Way dwarf galaxies. 
Fig.~\ref{fig:lambdaBias} shows how $\langle \mu(\lambda) \rangle \pm s$ with $s^2=  \langle \mu(\lambda)^2 \rangle -  \langle \mu(\lambda) \rangle^2$ evolves as a function of $\lambda$, as well as $\langle \mu_\gamma(\lambda) \rangle$ where we marginalize over $R_\mathrm{h}/\rmax$ alone, separately for cuspy ($\gamma=1$) and cored ($\gamma = 0$) DM profiles.
At $\lambda_\mathrm{eq}$, there is no functional dependence on $\gamma$ as $\langle \mu_{\gamma=0} \rangle=\langle \mu_{\gamma=1} \rangle \equiv \mu_\mathrm{eq}$.
For each DM profile, $\bar \lambda_{\gamma}$ minimizes the squared relative error $s_\gamma^2/\mu_\gamma^2$ when marginalizing over $R_\mathrm{h}/\rmax$ alone. 
The values $\bar \lambda=1.8,\,\bar \mu=3.5$ minimize the squared relative error after marginalizing over both $\gamma$ and $R_\mathrm{h}/\rmax$\footnote{
The value \mbox{$\tilde \mu = \langle \mu^2 \rangle / \langle \mu \rangle$} minimizes the squared relative error with \mbox{$ \langle (\mu/\tilde \mu -1)^2\rangle = s^2 / \langle \mu^2 \rangle < s^2 / \langle \mu \rangle^2 =  \langle (\mu/\bar \mu -1)^2\rangle $ }, however for this choice of $\tilde \mu$, the relative error equals \mbox{$\langle \mu / \tilde \mu -1 \rangle = - s^2 / \langle \mu^2 \rangle < 0$}, i.e. the estimator which minimizes the  squared relative error is biased.}. Table~\ref{tab:mu_lambda_parameters} lists the numerical values for $\mu,\lambda$.
For $\bar \mu, \bar \lambda$ as above, our minimum variance estimator becomes:
\begin{equation}
\label{eq:minvariance}
 M_\mathrm{est}(<1.8\,\Rh) \approx 3.5 \times 1.8 \, \Rh \, G^{-1} \, \langle \sigma_\mathrm{los}^2 \rangle ~.
\end{equation}
This $\mu=$const minimum variance estimator (equation~\ref{eq:minvariance}) allows to constrain the enclosed mass $M(<1.8\, \Rh)$ with an uncertainty of $s/ \bar \mu \approx 10$ per cent without making assumptions about the central slope of the underlying DM profile.

\begin{table}

\centering
\caption{Numerical values for $\lambda$ and $\mu$ of equation~\ref{eq:massEst}. $\bar \lambda$ minimizes the squared relative error $s^2/\mu^2$ when marginalizing both over $R_\mathrm{h}/\rmax$ and $\gamma$ (equation~\ref{eq:int}). When marginalizing instead only over $R_\mathrm{h}/\rmax$, then $\lambda_{\gamma,\mathrm{min}}$ minimizes the squared relative error for a given $\gamma$, whereas for $\lambda_{\mathrm{eq}}$, $\langle \mu_{\gamma=0} \rangle = \langle \mu_{\gamma=1} \rangle \equiv \mu_\mathrm{eq}$.}
\label{tab:mu_lambda_parameters}
\begin{tabular}{p{0.6cm}p{0.3cm}p{0.3cm}p{0.3cm}@{\hspace{0.7cm}}p{0.3cm}p{0.3cm}p{0.3cm}@{\hspace{0.7cm}}p{0.4cm}p{0.4cm}p{0.4cm}}
\toprule
           & $\bar \lambda$          & $\bar \mu$              & $\pm~s$                             & $\lambda_\mathrm{eq}$              & $\mu_\mathrm{eq}$                  & $\pm~s_\gamma$    & $\bar \lambda_{\gamma}$         & $\bar \mu_{\gamma}$         & $\pm~s_\gamma$  \\ 
           
\cmidrule(r{0.4cm}l{0.2cm}){2-4} \cmidrule(r{0.4cm}){5-7} \cmidrule(r){8-10}

$\gamma=0$ & \multirow{2}{*}{$1.83$} & \multirow{2}{*}{$3.51$} & \multirow{2}{*}{$0.33$}             & \multirow{2}{*}{$1.43$} & \multirow{2}{*}{$3.30$}                       & $0.61$ &              $1.88$                            &$3.57$                        & $0.48$ \\
$\gamma=1$ &                         &                         &                                     &                                    &                                    &$0.26$ &               $1.75$                            &$3.42$                        & $0.17$ \\
\bottomrule
\end{tabular}
\end{table}

The numerical values of $\bar \lambda, \bar \mu$ depend on the choice of shape for the DM profile (in our case $\alpha=1, \beta=5$ and $0\leq \gamma \leq 1$) and stellar profile (in our case Plummer spheres), as well as on the range of segregations $\Rh/\rmax$ and central slopes $\gamma$ to marginalize over. { Both $\bar \lambda_{\gamma}$ and $\bar \mu_{\gamma}$ differ by less than ten per cent between $\gamma=0$ and $\gamma=1$, i.e. the marginalization is fairly insensitive to the slope $\gamma$ for the range of segregations chosen. Recall from Figure \ref{fig:muLambda1} that for highly segregated systems, \mbox{$\mu_0(\lambda,\Rh/\rmax\rightarrow0)\rightarrow0$}, whereas \mbox{$\mu_1(\lambda,\Rh/\rmax\rightarrow 0) \rightarrow 3\lambda/2$}. The larger variation of $\mu_\gamma(\lambda,\Rh/\rmax)$ over the range of segregations for cored systems with respect to cuspy ones is reflected in the larger uncertainty $s_{\gamma=0} / \bar \mu_{\gamma=0} = 0.13 $ compared to $s_{\gamma=1} / \bar \mu_{\gamma=1} = 0.05$ }

Other studies have found similar values for $\lambda, \mu$ using conceptually different approaches: \citet{Walker2009} use Jeans analysis to motivate the choice of $\lambda = 1,~\mu = 5/2$ under the assumption of isotropy and a constant projected velocity dispersion. \citet{Wolf2010} use Jeans analysis to estimate the mass enclosed within the 3D half-light radius using the approximation $r_\mathrm{h} = 4/3 R_\mathrm{h}$, which translates to $\lambda = 4/3, \mu \approx 3$.
\citet{Amorisco2011, Amorisco2012} use a distribution-function approach to determine $\lambda \approx 1.67,~\mu \approx 3.50$ for Michie-King models, measuring $\sigma^2_\mathrm{los}$ at the projected half-light radius.
\citet{Campbell2016} find $\lambda \approx 1.77, \mu \approx 3.38$ empirically from a dispersion-supported sample of galaxies of the \textsc{apostle} simulation project \citep{SawalaFrenk2016}, averaging $\sigma^2_\mathrm{los}$ within $1.04 R_\mathrm{h}$.

\subsection{Method comparison using \emph{N}-body models}
\label{sec:methodcompenclosed}
In the following, we will test the minimum variance estimator on a suite of $N$-body models taken from the controlled cosmological simulations introduced in \cite{EPLG17}\footnote{In contrast to the set-up used in \cite{EPLG17}, we use $10^7$ $N$-body particles per satellite, the spatial resolution of the highest-resolving grid of the particle mesh code equals $\rmax/128$, and the \citet{miyamoto1975} galactic disc has mass and scales evolving with redshift as in \citet{Bullock2005}, with $M = 0.025\,M_{200}, a = \unit[3.5]{kpc}, b = \unit[0.3]{kpc}$ at $z=0$, where by $M_{200}$ we denote the virial mass of the Aquarius Aq-A2 main halo, and by $a,b$ the horizontal and vertical disc scale lengths, respectively.}.
This suite of $N$-body models consists of re-simulations of the accretion of all subhaloes with $M>\unit[10^8]{M_{\odot}}$ of the Aq-A2 merger tree \citep{Springel2008} on to an evolving, analytical Milky Way-like host potential consisting of an NFW halo and an axisymmetric disc. Subhaloes are modelled as equilibrium $N$-body realizations of either cuspy or cored Dehnen profiles with $10^7$ particles and injected in the host potential at $z_\mathrm{infall}$, defined by the peak of their mass evolution. All subhaloes are re-simulated with the same number of $N$-body particles independent of their mass, and the spatial resolution of the particle-mesh code is chosen as a function of the scale radius of each subhalo. This re-simulation technique allows us to follow the dynamical evolution of accreted substructures with the same numerical resolution for systems spanning many orders of magnitude in mass and size, and limits the impact of artificial disruption as discussed by \citet{vdb2018}. The subhaloes of the Aquarius simulation are cuspy, and have mass distributions before accretion consistent with the \citet{nfw1997} profile. We generate our cuspy and cored Dehnen equilibrium subhaloes keeping $M_{200}$ as given by the Aquarius simulation. We obtain $r_{-2}$ (i.e. the radius where $\mathrm{d} \ln \varrho / \mathrm{d} \ln r = -2$), which coincides with the NFW scale radius, from $M_{200}$ using the median \citet{prada2012} relation, and generate Dehnen models with that value of $r_{-2}$.
We model the evolution of the stellar populations under the assumption that they are collision-less and mass-less systems which only trace the underlying DM potential by assigning mass-to-light ratios to each $N$-body particle at infall, following the approach of \citet{Bullock2005}. At $z=0$, average observational properties of the stellar population, e.g. the velocity dispersion, can be obtained from the DM $N$-body particles by applying the initially assigned mass-to-light ratios as weights. We consider all haloes which at $z=0$ host a stellar population in virial equilibrium, requiring that $|2K+W|/(2K-W) < 0.2$ for the DM enclosed within $\rmax$, and $|2K_\mathrm{los}+W_\mathrm{los}|/(2K_\mathrm{los}-W_\mathrm{los}) < 0.05$ for the stars. Masses and velocity dispersions are measured from these $N$-body haloes using bound particles only.

\begin{figure}
\centering
  \includegraphics[height=7cm]{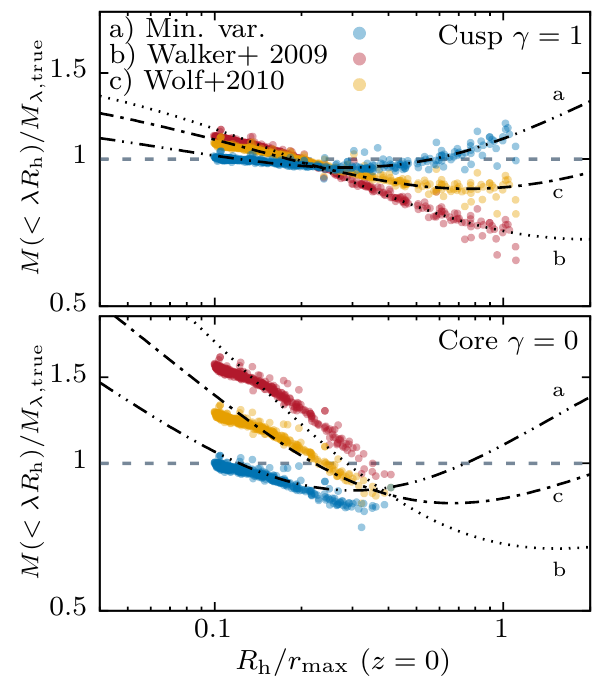}\includegraphics[width=7cm,angle=-90,origin=rb]{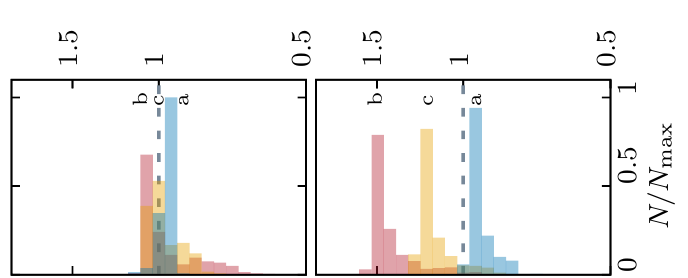}
  \caption{Estimated enclosed mass $M(<\lambda \Rh)$ at $z=0$ of $N$-body models taken from controlled re-simulations of the Aquarius A2 merger tree (see text), normalized by the true enclosed mass $M_\mathrm{\lambda,true}$ measured directly from the simulation, as a function of segregation $\Rh/\rmax$. The embedded stellar tracers have a segregation of $R_\mathrm{h,0}/r_\mathrm{max,0}=1/10$ at infall. We compare our minimum variance estimator ($\mu=3.5,\lambda=1.8$) against the estimators of \citet{Walker2009} and \citet{Wolf2010} separately for cuspy (top panel) and cored (bottom panel) DM haloes. Dotted lines show the theoretical prediction of $M(<\lambda \Rh)/M_\mathrm{\lambda,true}$ from the virial theorem { for stellar Plummer spheres embedded in $\{\alpha,\beta,\gamma\}=\{1,5,\gamma\}$ DM halo profiles.} The panels on the right are histograms obtained when summing over all haloes of the simulation.}
  \label{fig:menclosedmocks}
\end{figure}
Fig.~\ref{fig:menclosedmocks} shows the estimated enclosed masses $M(<\lambda \Rh)$ of $N$-body models at $z=0$ normalized by the true enclosed mass measured directly from the simulation as a function of segregation. The stellar population has a segregation of $R_\mathrm{h,0}/r_\mathrm{max,0} = 1/10$ at infall, which increases due to tides, spreading at $z=0$ over an interval $1/10 \lesssim \Rh/\rmax \lesssim 1$ for cuspy, and $1/10 \lesssim \Rh/\rmax \lesssim 1/2$ for cored systems. In the top panel, we compare our minimum variance estimator ($\mu=3.5,\lambda=1.8$) against the estimators of \citet{Walker2009} and \citet{Wolf2010} for cuspy DM haloes, whereas cored ones are plotted in the bottom panel. Dotted lines show the theoretical prediction of $M(<\lambda \Rh)/M_\mathrm{\lambda,true}$ from the virial theorem, i.e. $\mu_\mathrm{const}/\mu(\lambda, \Rh/\rmax)$ { for stellar Plummer spheres embedded in $\{\alpha,\beta,\gamma\}=\{1,5,\gamma\}$ DM halo profiles.}. The estimated masses follow closely the predictions of the virial theorem, which leads us to the conclusion that the main uncertainty on the estimated masses for a given DM profile shape originates from our ignorance on how deeply embedded the stellar population is within the DM halo.
{  The scatter around the predicted curve is mainly caused by deviations of the DM halo shape from the assumed $\{1,5,\gamma\}$ profile. }
For the range of segregations shown, when marginalizing over the entire population of $N$-body mocks, the masses estimated from the minimum variance estimator are within $\approx10$ per cent of the true value. The other two estimators in question give a similar accuracy for cuspy systems. For cored systems, the \citet{Wolf2010} estimator overestimates masses of cored systems by $\approx20$ per cent, whereas the \citet{Walker2009} estimator overestimates the masses of cored systems by $\approx50$ per cent. The uncertainties we find when estimating masses of cored haloes using $\mu=\mathrm{const}$ mass estimators are notably larger than the uncertainties described in \citet{Campbell2016}, derived from the \textsc{apostle} simulations \citep{SawalaFrenk2016}, and \citet{GonzalesSamaniego2017}, based on the \textsc{fire} simulations \citep{HopkinsFire2014}. This is due to the fact the $N$-body haloes in these simulations follow mass profiles close to the cuspy \citet{nfw1997} profile \citep[for \textsc{fire}, see also][discussing the core sizes as function of halo mass]{Chan2015Fire}, whereas our cored haloes have a large core size equal to the DM scale radius.

\section{Estimating the (total) halo mass}
\label{sec:totalmass}
In the previous sections, we focused on determining the mass enclosed within multiples of the half-light radius of dwarf galaxies embedded in DM haloes using $\mu =$const mass estimators (equation~\ref{eq:massEst}).
These estimators do not require to assume a specific DM profile when applying them to observational data - although the values of $\mu,\lambda$ can be motivated by a specific choice of profile, as done for the minimum-variance estimator, equation~\ref{eq:minvariance}. We now focus on estimating the total halo mass from measurements of the stellar velocity dispersion $\langle \sigma_\mathrm{los}^2 \rangle$ and the half-light radius $\Rh$.

\subsection{Degeneracy of halo structural parameters}
For a DM halo density profile $\varrho(r)$ with two free parameters, e.g. the total mass $M_\mathrm{tot}$ and a scale radius $a$ for the profile of equation~\ref{eq:betagammaprofile} with fixed $\{\alpha,\beta,\gamma\}$, the measurement of $\langle \sigma_\mathrm{los}^2 \rangle$ and half-light radius $\Rh$ can be used to constrain one parameter as a function of the other, but is not sufficient to determine numerical values for both parameters simultaneously. 
Let us explore this degeneracy for the case of $\{\alpha,\beta,\gamma\}=\{1,5,\gamma\}$ DM density profiles, i.e. DM profiles of tidally stripped systems, separately for the case of central DM cusps ($\gamma=1$) and cores ($\gamma=0$).

\subsubsection{Extrapolating the enclosed mass}
One way to obtain the degeneracy curve between the two halo structural parameters is to extrapolate the enclosed mass $M_\mathrm{est}(<\lambda\Rh)=\mu \lambda  \Rh \langle \sigma_\mathrm{los}^2 \rangle /G$ estimated using a $\mu$=const mass estimator. 
The cumulative mass profile corresponding to the density profile of equation~\ref{eq:betagammaprofile} with $\alpha=1,\beta=5$ reads
\begin{equation}
\label{eq:cumulativemassbetagamma}
 M(<r/a,\gamma)= M_\mathrm{tot} \left(  \frac{r}{a} + 4 -\gamma  \right)  \left( \frac{r}{a} \right)^{{3-\gamma}}  \left( \frac{r}{a} + 1 \right)^{{\gamma-4}}   ~.
\end{equation}
Defining $\tilde M(<r/a,\gamma) = M(<r/a,\gamma) / M_\mathrm{tot}$, which is a function of the radius $r$ and the scale radius $a$ in the combination $r/a$ alone, the total mass extrapolated from the enclosed mass $M_\mathrm{est}(<\lambda\Rh)$ then becomes
\begin{equation}
\label{eq:extrapM}
{M_\mathrm{tot}}(\Rh/a,\gamma) = {M_\mathrm{est}(<\lambda\Rh/a)}~{\tilde M^{-1}(<\lambda \Rh/a,\gamma)} ~.
\end{equation}
This total mass estimate is a function of the scale radius $a$ of the DM halo, or equivalently of $\rmax$, using 
\begin{equation}
\rmax = \frac{a}{2} \left( \sqrt{ 5 \gamma^2 - 34 \gamma + 57}  +\gamma -5 \right)~.
\end{equation}
Equation~\ref{eq:extrapM} is defined for a specific DM profile shape and, for a given $\gamma$, can be written as a function of the segregation parameter $\Rh/\rmax$ alone, but relies on a $\mu=$const mass estimate originally constructed to avoid the segregation dependence and the choice of a DM profile shape in the first place. Moreover, equation~\ref{eq:mu-general-form} tells us how to calculate $\mu$ directly from the virial theorem as a function of the segregation parameter and the choice of DM profile. This motivates to estimate the total mass directly from the virial theorem.

\subsubsection{Total mass from the virial theorem} 
Let's consider the projected potential energy term $W_\mathrm{los}$ of the spherical virial theorem, equation~\ref{eq:virial}. This term can be computed for given density profiles of the DM halo and the stellar tracer population. For $\{\alpha,\beta,\gamma\}=\{1,5,\gamma\}$ profiles with convergent total mass $M_\mathrm{tot}$, we now define $\tilde W_\mathrm{los}(\gamma, R_\mathrm{h}/\rmax) = W_\mathrm{los}(M, \gamma, R_\mathrm{h}/\rmax)/M_\mathrm{tot}$, which does depend on the scale $\rmax$ and inner slope $\gamma$ but not on the total mass $M_\mathrm{tot}$ of the DM halo. Then, using equation~\ref{eq:virial}, we find
\begin{equation}
\label{eq:virialM}
{M_\mathrm{tot}}(\Rh/\rmax,\gamma) = - { \langle \sigma_\mathrm{los}^2 \rangle }~{ \tilde W^{-1}_\mathrm{los}(\gamma, R_\mathrm{h}/\rmax) } ~.
\end{equation}
For given observed values of $\langle \sigma^2_\mathrm{los} \rangle$ and $R_\mathrm{h}$, and a choice of $\gamma$, the above equation~for $M_\mathrm{tot}$ is a function of the segregation parameter $\Rh/\rmax$ alone.

\subsubsection{$\{\rmax,\vmax\}$ curves} 
\begin{figure} 	 
  \centering
  \includegraphics[width=8.5cm]{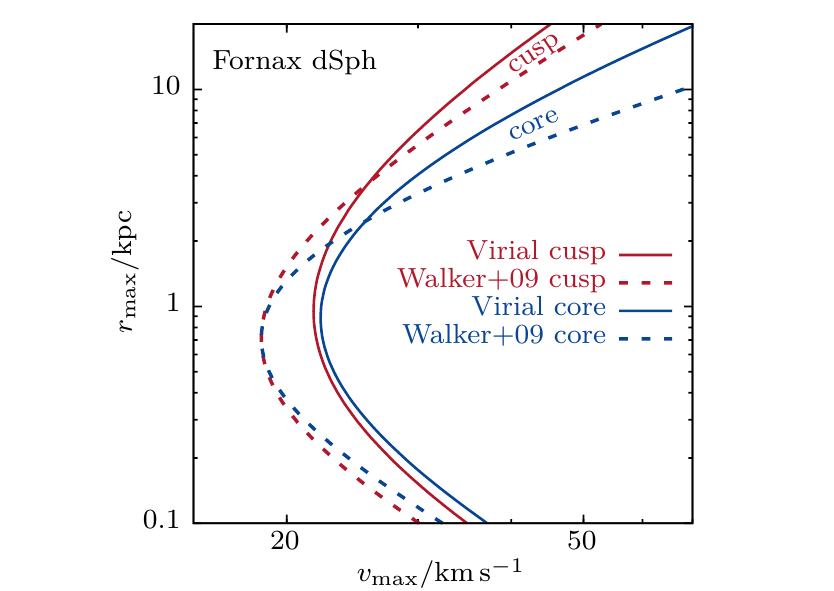}
  \caption{$\{\rmax,\vmax\}$ degeneracy curves for $\{\alpha,\beta,\gamma\}=\{1,5,\gamma\}$ DM haloes {compatible with the measurements of $\sqrt{\langle\sigma^2_\mathrm{los}\rangle}=(9.2 \pm 1.1)\,\unit[]{km\,s^{-1}}$ and $\Rh=(0.71\pm0.08)\,\unit[]{kpc}$ of the Fornax dSph (for references see Table \ref{tab:results}}). Curves obtained directly from the virial theorem are compared to those based on mass estimates using the \citet{Walker2009} estimator, shown separately for cuspy ($\gamma=1$) and cored ($\gamma=0$) DM profiles. }
  \label{fig:WalkerVsVirial}
\end{figure}

To facilitate comparison with literature, this mass-segregation degeneracy can be cast as a one-dimensional degeneracy curve in the plane of the structural parameters $\{\rmax,\vmax\}$ of the DM halo, as first discussed by \citet{PenarrubiaMcConnachieNavarro2008} for the case of stellar King profiles embedded in NFW haloes. For our case of tidally stripped systems with $\{\alpha,\beta,\gamma\}=\{1,5,\gamma\}$ profiles,
$\vmax = \left[G M(<\rmax)/{\rmax}\right]^{1/2}$ is calculated from equation~\ref{eq:cumulativemassbetagamma} using for the total mass either equation~\ref{eq:extrapM} ($\mu$=const extrapolation) or equation~\ref{eq:virialM} (virial theorem). 
When including observational errors, this degeneracy curve becomes equivalent to what \citet{AgnelloEvans2012} named \emph{virial stripe}.
Fig.~\ref{fig:WalkerVsVirial} compares the $\{\rmax,\vmax\}$ curves obtained directly from the virial theorem to the ones based on extrapolated masses using the \citet{Walker2009} estimator, i.e. $\lambda=1$ and using a constant $\mu=5/2$ in equation~\ref{eq:massEst}. {The curves are shown exemplarily for the Fornax dSph with a single stellar population\footnote{More complete models have been discussed in the literature: \citet{Walker2011} distinguish two and \citet{AmoriscoAgnelloEvans2013} three distinct stellar populations in the Fornax dSph.} modelled as Plummer sphere with projected half-light radius $\Rh=(0.71\pm0.08)\,\unit[]{kpc}$ and velocity dispersion $\sqrt{\langle\sigma^2_\mathrm{los}\rangle}=(9.2 \pm 1.1)\,\unit[]{km\,s^{-1}}$ (for references see Table \ref{tab:results}) embedded in an $\{\alpha,\beta,\gamma\}=\{1,5,\gamma\}$ DM halo.}
For low (high) values of $\rmax$, the virial theorem predicts larger (lower) values of $\vmax$ compatible with the same observed $\langle \sigma^2_\mathrm{los} \rangle,~ R_\mathrm{h}$. For the range of $\rmax$ shown in Fig.~\ref{fig:WalkerVsVirial}, the difference in the compatible values of $\vmax$ for the Fornax dwarf between the \citet{Walker2009} and the virial theorem is less than a factor unity. 

\subsection{Breaking the degeneracy: halo parameters from controlled cosmological simulations}
\begin{figure} 	 
  \centering
\includegraphics[width=8.5cm]{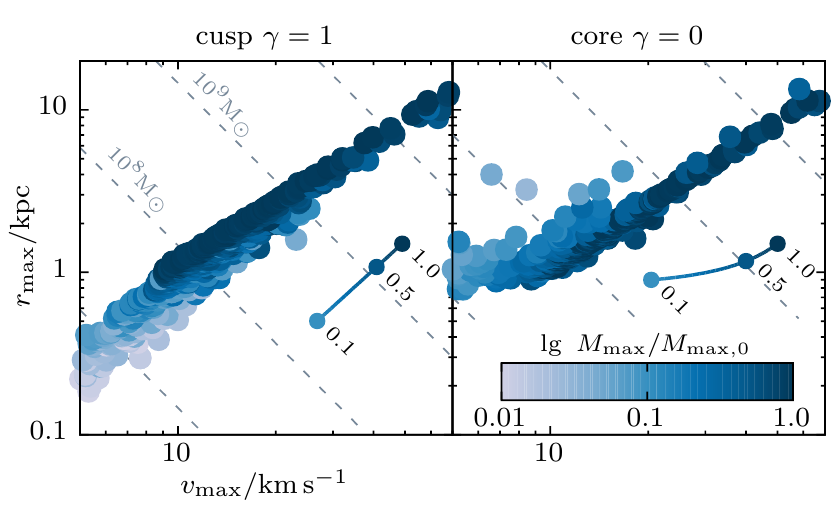}
\caption{$\{\rmax,\vmax\}$ values of DM subhaloes at $z=0$ of our controlled re-simulations of the Aquarius A2 merger tree. We model the accreted subhaloes to have either cuspy (left panel) or a cored (right panel) density profiles. The colour coding shows the mass $M_\mathrm{max}$ enclosed within $\mathrm{\rmax}$ at $z=0$ relative to the enclosed mass at infall. The solid curve and filled circles show the evolution within the $\{\rmax,\vmax\}$ plane along tidal evolutionary tracks (see Appendix \ref{appendix:tidal}), numbers denoting the fraction of remnant mass $M_\mathrm{max}$ enclosed within $\rmax$.}
\label{fig:CCCP2VmaxRmax}
\end{figure}

{
We aim to constrain the location of a DM halo hosting an observed dwarf galaxy along the $\{\rmax,\vmax\}$ degeneracy curve which is determined by the measured values of $\langle \sigma^2_\mathrm{los} \rangle$ and $\Rh$.
We use controlled simulations of the formation of Milky Way-like DM haloes to model the population of subhaloes at $z=0$.
For each Milky Way dwarf of Table \ref{tab:results} we infer the values of $\{\rmax,\vmax\}$ consistent with the observed kinematics and the population of simulated subhaloes of our controlled simulations discussed in section \ref{sec:methodcompenclosed}.
This allows us to put constraints on structural parameters like sizes and total masses of the DM haloes which host the Milky Way dwarfs for different assumptions about the population of DM subhaloes in the Milky Way. In specific we will study separately the cases in which the Milky Way DM halo population consists of DM subhaloes which were either cuspy or cored at accretion. 
}
This approach of breaking the $\{\rmax,\vmax\}$ degeneracy is closely related to the method introduced by \citet{PenarrubiaMcConnachieNavarro2008}, where a cosmological $\{\rmax,\vmax\}$ relation for field haloes was used. Using a relation for field haloes neglects the effects of tidal stripping on the DM profiles as experienced by satellite galaxies like the Milky Way dSphs, which we do take into account by using $\{\rmax,\vmax\}$ values of simulated tidally stripped haloes.
Fig.~\ref{fig:CCCP2VmaxRmax} shows the values of $\{\rmax,\vmax\}$ of the surviving population of simulated haloes at $z=0$, separately for cuspy (left-hand panel) and cored (right-hand panel) models. We only consider simulated haloes
where both the DM enclosed within $\rmax$ and the stars are approximately in virial equilibrium, requiring $|2K+W|/(2K-W) < 0.2$ for the DM, and $|2K_\mathrm{los}+W_\mathrm{los}|/(2K_\mathrm{los}-W_\mathrm{los}) < 0.05$ for the stars. When determining $\rmax$ and $\vmax$ of the $N$-body subhaloes, we consider bound particles only. The difference between cuspy and cored models becomes more pronounced with decreasing $\vmax$. Cored models are more extended than their cuspy counterparts for $\vmax \lesssim \unit[20]{km/s}$, and the relation shows substantially more scatter in the cored case. 
Also shown is the evolution of $\{\rmax,\vmax\}$ along \emph{tidal tracks} (see Appendix \ref{appendix:tidal}) as a function of the fraction of remnant mass $M_\mathrm{max}$ enclosed within $\rmax$ with respect to the value at infall. Whereas cuspy systems evolve approximately along the initial $\{\rmax,\vmax\}$ relation, keeping the range of $\{\rmax,\vmax\}$ values at a given total halo mass rather narrow, cored systems evolve away from the initial relation, causing the larger observed scatter.
While cored substructures are prone to disruption by tides, which explains the absence of systems with mass losses as high as for some cuspy models (see colour coding in Fig.~\ref{fig:CCCP2VmaxRmax}), in this study, we do not match abundances of the simulated systems but structural parameters only, either by fitting the enclosed mass $M(<\lambda \Rh)$ using a $\mu=$const mass estimators, or by fitting the observed velocity dispersion $\langle \sigma^2_\mathrm{los} \rangle$.

\subsubsection{Fitting the enclosed mass $M(<\lambda \Rh)$} When obtaining the $\{\rmax,\vmax\}$ degeneracy curve by extrapolating masses estimated through a $\mu=$const mass estimator (see equation~\ref{eq:extrapM}), it is a natural choice to break the degeneracy by selecting the simulated halo which matches best the estimated enclosed mass $M_\mathrm{est}(<\lambda \Rh)$ computed from the observed values of $\langle \sigma^2_\mathrm{los} \rangle$ and $\Rh$. For each simulated halo, we measure $\{\rmax,\vmax\}$ and compute the mass $M_\mathrm{sim}(<\lambda \Rh)$ assuming an $\{\alpha,\beta,\gamma\}=\{1,5,\gamma\}$ profile with structural parameters $\{\rmax,\vmax\}$.
We then select the halo which minimizes
\begin{equation}
 \label{equ:medianChi2mass}
 \chi^2_{M(<\lambda \Rh)} =  {\Big( M_\mathrm{est}(<\lambda \Rh) - M_\mathrm{sim}(<\lambda \Rh)  \Big)^2 } ~/~ \mathrm{var}~,
\end{equation}
where by $\mathrm{var}$ we denote the propagated variances from the observational uncertainties on $\langle \sigma^2_\mathrm{los} \rangle$ and $R_\mathrm{h}$. The variance is estimated through Monte-Carlo runs assuming Gaussian errors on $\langle \sigma^2_\mathrm{los} \rangle$ and $R_\mathrm{h}$ to account for the correlation between var$(M_\mathrm{est}(<\lambda \Rh))$ and var$(M_\mathrm{sim}(<\lambda \Rh))$  introduced by the uncertainty in $\Rh$. 

\subsubsection{Fitting the velocity dispersion $\langle \sigma^2_\mathrm{los} \rangle$}
When obtaining the degeneracy curve directly from the viral theorem (equation~\ref{eq:virialM}), for each simulated halo with structural parameters $\{\rmax,\vmax\}$, we calculate the expected velocity dispersion $\langle \sigma^2_\mathrm{los,sim} \rangle$ from the virial theorem, using the observed value of $\Rh$,
\begin{equation}
\label{eq:directsigma2fit}
\langle \sigma^2_\mathrm{los,sim} \rangle = - W_\mathrm{los} = \frac{4 \pi G}{3} \int_0^\infty r \nu_\star(r) M(<r) \,\mathrm{d}r ~.
\end{equation}
In the above equation, $\nu_\star(r)$ is the density profile of a Plummer sphere, $\{\alpha_\star,\beta_\star,\gamma_\star\} = \{2,5,0\}$, with projected half-light radius $\Rh$, and $M(<r)$ is the cumulative mass profile for an $\{\alpha,\beta,\gamma\}=\{1,5,\gamma\}$ DM halo with structural parameters $\{\rmax,\vmax\}$.
We then confront this expected velocity dispersion $\langle \sigma^2_\mathrm{los,sim} \rangle$ to the observed velocity dispersion $\langle \sigma^2_\mathrm{los} \rangle$ by computing
\begin{equation}
 \label{equ:medianChi2sigma}
 \chi^2_{\langle \sigma^2_\mathrm{los} \rangle} =  {\left( \langle \sigma^2_\mathrm{los} \rangle - \langle \sigma^2_\mathrm{los,sim} \rangle  \right)^2 }~/~\mathrm{var} ~.
\end{equation}
The variance term in the denominator is again obtained by Monte-Carlo sampling, assuming Gaussian errors on the observables $\Rh$ and $\langle \sigma^2_\mathrm{los} \rangle$ .

\subsection{Test of self-consistency using \emph{N}-body models}
\begin{figure} 	 
  \centering
  \includegraphics[width=\columnwidth]{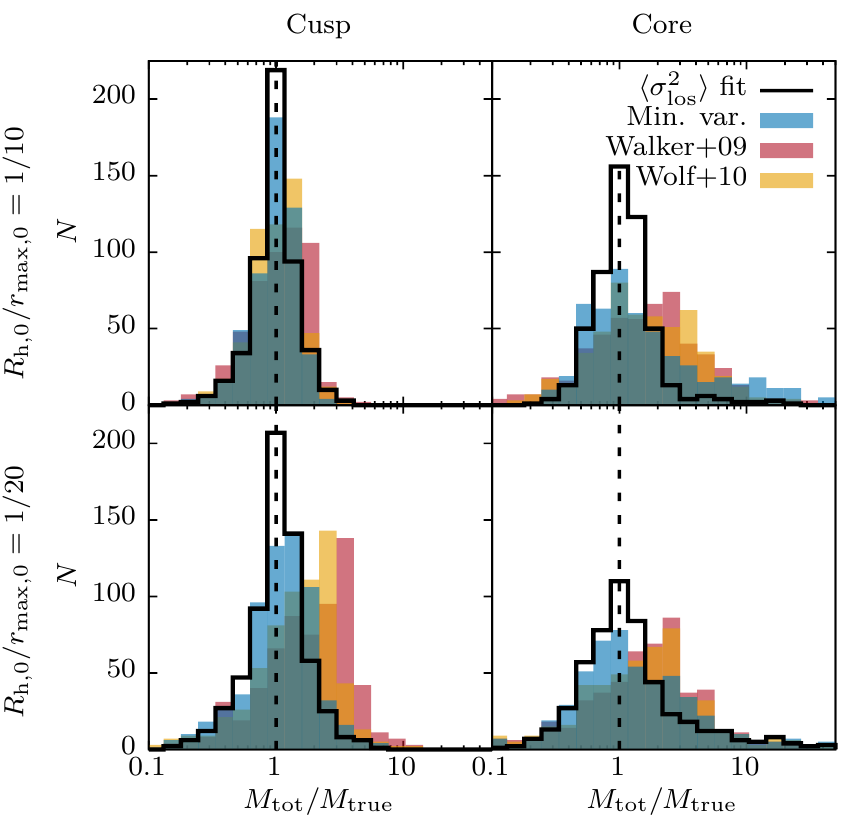} 
     \caption{Histograms of estimated over true total halo mass $M_\mathrm{tot}/M_\mathrm{true}$ for four different catalogues of $N$-body subhaloes obtained from re-simulations of the Aquarius A2 merger tree: cuspy (cored) models are shown in the left (right) column, and the embedded stellar populations have an initial segregation of $R_\mathrm{h,0}/r_\mathrm{max,0} = 1/10~(1/20)$ in the top row (bottom row). For each catalogue of $N$-body subhaloes, four different methods are used to estimate the total halo mass: direct fits of $\langle \sigma^2_\mathrm{los} \rangle$ (see equation~\ref{eq:directsigma2fit}), as well as masses extrapolated from three $\mu=$const estimators, including the minimum variance estimator introduced in section \ref{sec:unbiased}.}
  \label{fig:FourHistograms}
\end{figure}  
Before applying the mass estimator to kinematic data of Milky Way dwarf galaxies in section \ref{sec:dwarf}, as a consistency check we recover the masses of the catalogue of $N$-body subhaloes described in section \ref{sec:methodcompenclosed}, i.e. we embed stellar populations in the same DM subhaloes which we use to break the $\{\rmax,\vmax\}$ degeneracy. In Fig.~\ref{fig:FourHistograms} we compare the masses estimated through direct $\langle \sigma^2_\mathrm{los} \rangle$ fits to masses extrapolated from $\mu=$const estimators, assuming DM profiles with $\alpha=1$ and $\beta=5$ (see equation~\ref{eq:betagammaprofile}). The logarithms of the masses obtained from the virial estimator (black lines) are distributed approximately symmetrically around the true mass with $\langle \declog\,M_\mathrm{tot}/M_\mathrm{true} \rangle = 0.0 \pm 0.2$ for cuspy systems, and $0.0\pm0.2$ ($0.1\pm0.5$) for cored systems with an initial segregation of $R_\mathrm{h,0}/r_\mathrm{max,0}=1/10$ $(1/20)$.

On average, masses extrapolated from the three $\mu=$const estimators are overestimated, i.e. $\langle \declog\,M_\mathrm{tot}/M_\mathrm{true} \rangle > 0$ for all four subhalo catalogues. For masses extrapolated for cuspy systems from the minimum variance $\mu=$const estimator introduced in section \ref{sec:unbiased} we find 
$\langle \declog\,M_\mathrm{tot}/M_\mathrm{true} \rangle \sim 0.0 \pm 0.3$, and $\lesssim 0.2 \pm 0.5$ for cored systems, i.e. bias and spread of the distribution are larger than for the masses obtained directly from the virial theorem. 
The bias $\langle \declog\,M_\mathrm{tot}/M_\mathrm{true} \rangle$ is larger for the other two $\mu=$const estimators. For systems with an initial segregation of 1/20, both
the \citet{Walker2009} and the \citet{Wolf2010} estimator give $\langle \declog\,M_\mathrm{tot}/M_\mathrm{true} \rangle = 0.2 \pm 0.3$ ($0.2 \pm 0.6$) for cuspy (cored) systems.
Overall, we find larger biases and spreads in the distribution of recovered masses for cored systems than for cuspy ones. Cored systems do lose mass to tides more easily than cuspy systems, and their mass profiles show more extra-tidal features, i.e. our assumption of $\{\alpha,\beta,\gamma\}=\{1,5,0\}$ profiles does not match cored systems well which have lost large fractions of their initial mass.

\section{Discussion: The halo masses of Milky Way dwarf spheroidal galaxies}
\label{sec:MWdwarfs} {
We apply the minimum variance mass estimator to observational data of Milky Way dwarf galaxies compiled by \citet{McConnachie2012} (supplemented with more recent measurements as detailed in Table \ref{tab:results})
to infer the structural properties of DM haloes compatible with the observed values of $\langle \sigma^2_\mathrm{los} \rangle$ and $R_\mathrm{h}$. }The velocity dispersions measured for Milky Way dSphs are often limited to one measurement per galaxy, e.g. the central velocity dispersion, or the velocity dispersion averaged within a certain radius, not necessarily up to infinity. Limited by the data available, for the following mass estimates we use the observed $\sigma_\mathrm{los,obs}$ as listed in \citet{McConnachie2012} and assume $\sigma_\mathrm{los,obs}^2 = \langle \sigma_\mathrm{los}^2 \rangle$ in the notation of equation~\ref{eq:pressureTerm}.

\afterpage{
 \begin{landscape}
  \begin{table}
\begin{threeparttable}
{
\centering
\caption{{Structural parameters for the brightest Milky Way dSphs. $V$-band luminosity $L_\star$, half-light radius $\Rh$ and line-of-sight velocity dispersion $\sigma_\mathrm{los}$ are taken from \citet{McConnachie2012} unless indicated differently. References to the number of member stars $N$ used for the computation of $\sigma_\mathrm{los}$ are given in the table. The mass $M_{1.8} \equiv M(<1.8\,\Rh)$ enclosed within $1.8$ half-light radii $\Rh$ is estimated using the minimum-variance estimator of equation~\ref{eq:minvariance}, which is then used to obtain the mean density $\varrho_{1.8} \equiv \langle \varrho(<1.8\,\Rh)\rangle$. Uncertainties on $M_{1.8}$ and $\varrho_{1.8}$ are obtained by Monte Carlo sampling symmetric Gaussian errors on $\sigma_\mathrm{los}$ and $\Rh$. 
Structural parameters of the DM haloes, i.e. $\rmax$, $\vmax$ and total halo mass $M_\mathrm{tot}$, are obtained by fitting the observed velocity dispersion $\sigma_\mathrm{los}$ to the $N$-body subhaloes of our re-simulations of the Aquarius A2 merger tree, as discussed in section \ref{sec:totalmass}. These quantities are listed separately for fits to $N$-body subhaloes with cuspy ($\gamma=1$) and cored ($\gamma=0$) DM profiles. Total halo masses of $N$-body subhaloes are extrapolated from the mass $M_\mathrm{max}$ enclosed within $\rmax$ under the assumption of $\{\alpha,\beta,\gamma\}=\{1,5,\gamma\}$-profiles. Uncertainties on $\rmax$, $\vmax$ and $M_\mathrm{tot}$ indicate the range of values of those $N$-body subhaloes which satisfy $\chi^2 \leq \chi^2_\mathrm{min} +1$.  
}}
\label{tab:results}
\renewcommand{\arraystretch}{1.47}
{
\begin{tabular}{@{\hspace{-0.4cm}}rr@{}c@{}lccr@{\hspace{0.9cm}} r@{}c@{}l @{\hspace{0.9cm}} r@{}c@{}l@{\hspace{1.1cm}} r@{}c@{}l r@{}c@{}l r@{}c@{}l c @{\hspace{0.9cm}} r@{}c@{}l r@{}c@{}l r@{}c@{}l c}
\toprule
 & &&&&&& &&&&&& \multicolumn{10}{c}{\textsc{Cusp$~~~~~~~~$}} & \multicolumn{10}{c}{\textsc{Core}} \\ \addlinespace
 & \multicolumn{3}{@{\hspace{-0.1cm}}c}{$\displaystyle\declog~\frac{L_\star}{\unit[]{L_{\odot}}}$} 
 & $\displaystyle \frac{R_\mathrm{h}}{\mathrm{kpc}}$ 
 & $\displaystyle\frac{\sigma_\mathrm{los}}{\kms}$
 & $N~$ & \multicolumn{3}{@{\hspace{-0.2cm}}l}{$\displaystyle \declog \frac{M_{1.8}}{\mathrm{M_{\odot}}}$} 
 & \multicolumn{3}{@{\hspace{-0.8cm}}c}{$\displaystyle \declog\,\frac{\varrho_{1.8}}{\mathrm{M_{\odot}} \mathrm{kpc^{-3}} }$} 
 & \multicolumn{3}{@{}c}{$\displaystyle\frac{\vmax}{\kms}$}  
 & \multicolumn{3}{c}{$\displaystyle\frac{\rmax}{\mathrm{kpc}}$}  
 & \multicolumn{3}{c}{$\displaystyle\declog~\frac{M_\mathrm{tot}}{\mathrm{M_{\odot}}}$}  
 &  $\chi^2_\mathrm{min}$ 
 & \multicolumn{3}{@{}c}{$\displaystyle\frac{\vmax}{\kms}$}  
 & \multicolumn{3}{c}{$\displaystyle\frac{\rmax}{\mathrm{kpc}}$}  
 & \multicolumn{3}{c}{$\displaystyle\declog~\frac{M_\mathrm{tot}}{ \mathrm{M_{\odot}}}$}  &  $\chi^2_\mathrm{min}$ \\
 \cmidrule(r{0.7cm}){2-7} \cmidrule(l{-0.2cm}r{0.6cm}){8-13} \cmidrule(r{0.85cm}){14-23} \cmidrule(l{-0.05cm}r{0.15cm}){24-33}
Sagittarius dSph & ${7.34}$ & ${}^{+}_{-}$ & ${}^{0.18}_{0.18}$ & $2.62 \pm 0.20$ & $11.40 \pm 0.70$ & 114 \tnotex{tn:1} & ${8.7}$ & ${}^{+}_{-}$ & ${}^{0.1}_{0.1}$ & ${6.1}$ & ${}^{+}_{-}$ & ${}^{0.1}_{0.1}$ & ${21.8}$ & ${}^{+}_{-}$ & ${}^{1.4}_{1.3}$ & ${2.0}$ & ${}^{+}_{-}$ & ${}^{1.1}_{0.4}$ & ${8.8}$ & ${}^{+}_{-}$ & ${}^{0.2}_{0.1}$ & $0.0$ & ${21.5}$ & ${}^{+}_{-}$ & ${}^{0.7}_{0.8}$ & ${2.9}$ & ${}^{+}_{-}$ & ${}^{0.1}_{0.4}$ & ${8.9}$ & ${}^{+}_{-}$ & ${}^{0.0}_{0.1}$ & $0.0$ \\
Fornax & ${7.31}$ & ${}^{+}_{-}$ & ${}^{0.19}_{0.20}$ & $0.71 \pm 0.08$ & $11.70 \pm 0.90$ & 2633 \tnotex{tn:2} & ${8.2}$ & ${}^{+}_{-}$ & ${}^{0.1}_{0.1}$ & ${7.2}$ & ${}^{+}_{-}$ & ${}^{0.1}_{0.1}$ & ${23.4}$ & ${}^{+}_{-}$ & ${}^{5.5}_{2.0}$ & ${2.3}$ & ${}^{+}_{-}$ & ${}^{2.2}_{0.7}$ & ${8.9}$ & ${}^{+}_{-}$ & ${}^{0.5}_{0.2}$ & $0.0$ & ${31.7}$ & ${}^{+}_{-}$ & ${}^{26.6}_{4.0}$ & ${4.6}$ & ${}^{+}_{-}$ & ${}^{8.8}_{0.6}$ & ${9.4}$ & ${}^{+}_{-}$ & ${}^{1.0}_{0.2}$ & $0.0$ \\
Leo I & ${6.74}$ & ${}^{+}_{-}$ & ${}^{0.17}_{0.18}$ & $0.25 \pm 0.03$ & $9.20 \pm 0.40$ \tnotex{tn:3} & 328 \tnotex{tn:3} & ${7.5}$ & ${}^{+}_{-}$ & ${}^{0.1}_{0.1}$ & ${7.9}$ & ${}^{+}_{-}$ & ${}^{0.1}_{0.1}$ & ${27.9}$ & ${}^{+}_{-}$ & ${}^{4.2}_{6.4}$ & ${3.7}$ & ${}^{+}_{-}$ & ${}^{1.3}_{1.8}$ & ${9.3}$ & ${}^{+}_{-}$ & ${}^{0.3}_{0.5}$ & $0.0$ & ${65.0}$ & ${}^{+}_{-}$ & ${}^{2.4}_{0.0}$ & ${10.7}$ & ${}^{+}_{-}$ & ${}^{0.6}_{0.0}$ & ${10.4}$ & ${}^{+}_{-}$ & ${}^{0.1}_{0.0}$ & $2.8$ \\
Sculptor & ${6.36}$ & ${}^{+}_{-}$ & ${}^{0.25}_{0.26}$ & $0.28 \pm 0.04$ & $9.20 \pm 1.10$ \tnotex{tn:2} & 1541 \tnotex{tn:2} & ${7.5}$ & ${}^{+}_{-}$ & ${}^{0.1}_{0.1}$ & ${7.8}$ & ${}^{+}_{-}$ & ${}^{0.2}_{0.2}$ & ${27.2}$ & ${}^{+}_{-}$ & ${}^{4.9}_{9.4}$ & ${3.9}$ & ${}^{+}_{-}$ & ${}^{1.1}_{2.5}$ & ${9.3}$ & ${}^{+}_{-}$ & ${}^{0.2}_{0.8}$ & $0.0$ & ${65.0}$ & ${}^{+}_{-}$ & ${}^{8.6}_{24.8}$ & ${10.7}$ & ${}^{+}_{-}$ & ${}^{3.7}_{4.6}$ & ${10.4}$ & ${}^{+}_{-}$ & ${}^{0.2}_{0.7}$ & $0.1$ \\
Leo II & ${5.87}$ & ${}^{+}_{-}$ & ${}^{0.17}_{0.18}$ & $0.18 \pm 0.04$ & $6.60 \pm 0.70$ & 171 \tnotex{tn:4} & ${7.0}$ & ${}^{+}_{-}$ & ${}^{0.1}_{0.2}$ & ${7.9}$ & ${}^{+}_{-}$ & ${}^{0.2}_{0.2}$ & ${16.9}$ & ${}^{+}_{-}$ & ${}^{9.9}_{3.5}$ & ${1.6}$ & ${}^{+}_{-}$ & ${}^{2.4}_{0.7}$ & ${8.5}$ & ${}^{+}_{-}$ & ${}^{0.8}_{0.4}$ & $0.0$ & ${65.0}$ & ${}^{+}_{-}$ & ${}^{8.6}_{46.8}$ & ${10.7}$ & ${}^{+}_{-}$ & ${}^{3.7}_{9.1}$ & ${10.4}$ & ${}^{+}_{-}$ & ${}^{0.2}_{1.9}$ & $0.1$ \\
Sextans (I) & ${5.64}$ & ${}^{+}_{-}$ & ${}^{0.24}_{0.24}$ & $0.69 \pm 0.04$ & $7.90 \pm 1.30$ & 947 \tnotex{tn:2} & ${7.8}$ & ${}^{+}_{-}$ & ${}^{0.1}_{0.2}$ & ${6.9}$ & ${}^{+}_{-}$ & ${}^{0.1}_{0.2}$ & ${15.5}$ & ${}^{+}_{-}$ & ${}^{3.0}_{3.4}$ & ${2.0}$ & ${}^{+}_{-}$ & ${}^{0.5}_{1.1}$ & ${8.5}$ & ${}^{+}_{-}$ & ${}^{0.3}_{0.5}$ & $0.0$ & ${17.2}$ & ${}^{+}_{-}$ & ${}^{3.8}_{4.5}$ & ${2.4}$ & ${}^{+}_{-}$ & ${}^{0.5}_{1.1}$ & ${8.6}$ & ${}^{+}_{-}$ & ${}^{0.3}_{0.5}$ & $0.0$ \\
Carina & ${5.58}$ & ${}^{+}_{-}$ & ${}^{0.25}_{0.26}$ & $0.25 \pm 0.04$ & $6.60 \pm 1.20$ & 1982 \tnotex{tn:2} & ${7.2}$ & ${}^{+}_{-}$ & ${}^{0.2}_{0.2}$ & ${7.6}$ & ${}^{+}_{-}$ & ${}^{0.2}_{0.2}$ & ${16.4}$ & ${}^{+}_{-}$ & ${}^{5.5}_{6.5}$ & ${2.1}$ & ${}^{+}_{-}$ & ${}^{1.1}_{1.6}$ & ${8.6}$ & ${}^{+}_{-}$ & ${}^{0.4}_{1.0}$ & $0.0$ & ${29.0}$ & ${}^{+}_{-}$ & ${}^{44.6}_{16.0}$ & ${4.2}$ & ${}^{+}_{-}$ & ${}^{10.2}_{3.0}$ & ${9.3}$ & ${}^{+}_{-}$ & ${}^{1.3}_{1.2}$ & $0.0$ \\
Ursa Minor & ${5.46}$ & ${}^{+}_{-}$ & ${}^{0.24}_{0.24}$ & $0.18 \pm 0.03$ & $8.10 \pm 1.10$ \tnotex{tn:5} & 235 \tnotex{tn:5} & ${7.2}$ & ${}^{+}_{-}$ & ${}^{0.1}_{0.1}$ & ${8.1}$ & ${}^{+}_{-}$ & ${}^{0.2}_{0.2}$ & ${23.4}$ & ${}^{+}_{-}$ & ${}^{16.3}_{5.7}$ & ${2.3}$ & ${}^{+}_{-}$ & ${}^{4.5}_{0.9}$ & ${8.9}$ & ${}^{+}_{-}$ & ${}^{0.9}_{0.5}$ & $0.0$ & ${65.0}$ & ${}^{+}_{-}$ & ${}^{6.1}_{24.8}$ & ${10.7}$ & ${}^{+}_{-}$ & ${}^{2.5}_{4.6}$ & ${10.4}$ & ${}^{+}_{-}$ & ${}^{0.2}_{0.7}$ & $1.6$ \\
Draco & ${5.46}$ & ${}^{+}_{-}$ & ${}^{0.14}_{0.15}$ & $0.22 \pm 0.02$ & $9.10 \pm 1.20$ & 413 \tnotex{tn:6} & ${7.4}$ & ${}^{+}_{-}$ & ${}^{0.1}_{0.1}$ & ${8.0}$ & ${}^{+}_{-}$ & ${}^{0.1}_{0.1}$ & ${32.1}$ & ${}^{+}_{-}$ & ${}^{7.7}_{11.9}$ & ${4.9}$ & ${}^{+}_{-}$ & ${}^{1.9}_{3.3}$ & ${9.5}$ & ${}^{+}_{-}$ & ${}^{0.3}_{0.8}$ & $0.0$ & ${65.0}$ & ${}^{+}_{-}$ & ${}^{6.1}_{16.7}$ & ${10.7}$ & ${}^{+}_{-}$ & ${}^{2.5}_{3.1}$ & ${10.4}$ & ${}^{+}_{-}$ & ${}^{0.2}_{0.4}$ & $1.6$ \\
Canes Venatici (I) & ${5.37}$ & ${}^{+}_{-}$ & ${}^{0.12}_{0.12}$ & $0.56 \pm 0.04$ & $7.60 \pm 0.40$ & 214 \tnotex{tn:7} & ${7.7}$ & ${}^{+}_{-}$ & ${}^{0.1}_{0.1}$ & ${7.0}$ & ${}^{+}_{-}$ & ${}^{0.1}_{0.1}$ & ${15.3}$ & ${}^{+}_{-}$ & ${}^{1.1}_{1.9}$ & ${1.9}$ & ${}^{+}_{-}$ & ${}^{0.3}_{0.9}$ & ${8.5}$ & ${}^{+}_{-}$ & ${}^{0.1}_{0.4}$ & $0.0$ & ${17.1}$ & ${}^{+}_{-}$ & ${}^{1.4}_{2.1}$ & ${2.2}$ & ${}^{+}_{-}$ & ${}^{0.5}_{0.6}$ & ${8.6}$ & ${}^{+}_{-}$ & ${}^{0.1}_{0.2}$ & $0.0$ \\
Crater 2 & ${5.21}$ & ${}^{+}_{-}$ & ${}^{0.04}_{0.04}$ \tnotex{tn:8} & $1.07 \pm 0.08$ \tnotex{tn:8} & $2.70 \pm 0.30$ \tnotex{tn:9} & 390 \tnotex{tn:9} & ${7.1}$ & ${}^{+}_{-}$ & ${}^{0.1}_{0.1}$ & ${5.6}$ & ${}^{+}_{-}$ & ${}^{0.1}_{0.1}$ & ${5.9}$ & ${}^{+}_{-}$ & ${}^{0.6}_{0.7}$ & ${0.4}$ & ${}^{+}_{-}$ & ${}^{0.0}_{0.1}$ & ${7.0}$ & ${}^{+}_{-}$ & ${}^{0.1}_{0.2}$ & $0.0$ & ${5.2}$ & ${}^{+}_{-}$ & ${}^{1.4}_{0.6}$ & ${0.9}$ & ${}^{+}_{-}$ & ${}^{3.1}_{0.1}$ & ${7.2}$ & ${}^{+}_{-}$ & ${}^{0.8}_{0.1}$ & $0.0$ \\
Leo T & ${5.14}$ & ${}^{+}_{-}$ & ${}^{0.24}_{0.24}$ & $0.12 \pm 0.01$ & $7.50 \pm 1.60$ & 19 \tnotex{tn:7} & ${7.0}$ & ${}^{+}_{-}$ & ${}^{0.2}_{0.2}$ & ${8.4}$ & ${}^{+}_{-}$ & ${}^{0.2}_{0.2}$ & ${37.5}$ & ${}^{+}_{-}$ & ${}^{16.6}_{22.4}$ & ${6.2}$ & ${}^{+}_{-}$ & ${}^{3.5}_{5.1}$ & ${9.8}$ & ${}^{+}_{-}$ & ${}^{0.5}_{1.5}$ & $0.0$ & ${18.2}$ & ${}^{+}_{-}$ & ${}^{55.4}_{5.3}$ & ${1.6}$ & ${}^{+}_{-}$ & ${}^{12.8}_{0.4}$ & ${8.5}$ & ${}^{+}_{-}$ & ${}^{2.2}_{0.4}$ & $1.9$ \\
Hercules & ${4.58}$ & ${}^{+}_{-}$ & ${}^{0.19}_{0.21}$ & $0.33 \pm 0.06$ & $3.72 \pm 0.91$ \tnotex{tn:10} & 28 \tnotex{tn:10} & ${6.8}$ & ${}^{+}_{-}$ & ${}^{0.2}_{0.3}$ & ${6.9}$ & ${}^{+}_{-}$ & ${}^{0.3}_{0.3}$ & ${6.9}$ & ${}^{+}_{-}$ & ${}^{2.0}_{1.8}$ & ${0.4}$ & ${}^{+}_{-}$ & ${}^{0.5}_{0.2}$ & ${7.1}$ & ${}^{+}_{-}$ & ${}^{0.6}_{0.6}$ & $0.0$ & ${8.1}$ & ${}^{+}_{-}$ & ${}^{8.5}_{2.7}$ & ${1.2}$ & ${}^{+}_{-}$ & ${}^{3.0}_{0.4}$ & ${7.6}$ & ${}^{+}_{-}$ & ${}^{1.2}_{0.5}$ & $0.0$ \\
Bootes (I) & ${4.46}$ & ${}^{+}_{-}$ & ${}^{0.11}_{0.11}$ & $0.24 \pm 0.02$ & $2.40 \pm 0.70$ & 50 \tnotex{tn:11} & ${6.3}$ & ${}^{+}_{-}$ & ${}^{0.2}_{0.3}$ & ${6.8}$ & ${}^{+}_{-}$ & ${}^{0.2}_{0.3}$ & ${4.4}$ & ${}^{+}_{-}$ & ${}^{1.3}_{1.1}$ & ${0.2}$ & ${}^{+}_{-}$ & ${}^{0.2}_{0.1}$ & ${6.4}$ & ${}^{+}_{-}$ & ${}^{0.5}_{0.4}$ & $0.0$ & ${5.7}$ & ${}^{+}_{-}$ & ${}^{2.8}_{2.3}$ & ${1.1}$ & ${}^{+}_{-}$ & ${}^{2.9}_{0.4}$ & ${7.3}$ & ${}^{+}_{-}$ & ${}^{0.8}_{0.6}$ & $0.0$ \\
Leo IV & ${4.27}$ & ${}^{+}_{-}$ & ${}^{0.19}_{0.20}$ & $0.21 \pm 0.04$ & $3.30 \pm 1.70$ & 18 \tnotex{tn:7} & ${6.5}$ & ${}^{+}_{-}$ & ${}^{0.4}_{0.6}$ & ${7.2}$ & ${}^{+}_{-}$ & ${}^{0.4}_{0.6}$ & ${6.4}$ & ${}^{+}_{-}$ & ${}^{4.6}_{5.0}$ & ${0.5}$ & ${}^{+}_{-}$ & ${}^{0.8}_{0.4}$ & ${7.2}$ & ${}^{+}_{-}$ & ${}^{0.9}_{1.7}$ & $0.0$ & ${10.2}$ & ${}^{+}_{-}$ & ${}^{8.2}_{7.6}$ & ${1.8}$ & ${}^{+}_{-}$ & ${}^{2.4}_{1.1}$ & ${8.0}$ & ${}^{+}_{-}$ & ${}^{0.8}_{1.6}$ & $0.0$ \\
Ursa Major (I) & ${4.15}$ & ${}^{+}_{-}$ & ${}^{0.16}_{0.16}$ & $0.32 \pm 0.05$ & $7.60 \pm 1.00$ & 39 \tnotex{tn:7} & ${7.4}$ & ${}^{+}_{-}$ & ${}^{0.1}_{0.1}$ & ${7.5}$ & ${}^{+}_{-}$ & ${}^{0.2}_{0.2}$ & ${16.4}$ & ${}^{+}_{-}$ & ${}^{7.1}_{3.2}$ & ${1.5}$ & ${}^{+}_{-}$ & ${}^{2.0}_{0.6}$ & ${8.5}$ & ${}^{+}_{-}$ & ${}^{0.7}_{0.4}$ & $0.0$ & ${24.7}$ & ${}^{+}_{-}$ & ${}^{48.9}_{8.0}$ & ${3.1}$ & ${}^{+}_{-}$ & ${}^{11.3}_{1.5}$ & ${9.0}$ & ${}^{+}_{-}$ & ${}^{1.6}_{0.6}$ & $0.0$ \\
Leo V & ${4.04}$ & ${}^{+}_{-}$ & ${}^{0.21}_{0.21}$ & $0.13 \pm 0.03$ & $3.70 \pm 1.85$ & 7 \tnotex{tn:12} & ${6.4}$ & ${}^{+}_{-}$ & ${}^{0.4}_{0.6}$ & ${7.7}$ & ${}^{+}_{-}$ & ${}^{0.4}_{0.6}$ & ${8.1}$ & ${}^{+}_{-}$ & ${}^{8.4}_{6.8}$ & ${0.7}$ & ${}^{+}_{-}$ & ${}^{1.5}_{0.6}$ & ${7.5}$ & ${}^{+}_{-}$ & ${}^{1.1}_{2.0}$ & $0.0$ & ${14.0}$ & ${}^{+}_{-}$ & ${}^{59.6}_{11.4}$ & ${1.7}$ & ${}^{+}_{-}$ & ${}^{12.7}_{1.0}$ & ${8.3}$ & ${}^{+}_{-}$ & ${}^{2.4}_{1.8}$ & $0.0$ \\
Canes Venatici II & ${3.90}$ & ${}^{+}_{-}$ & ${}^{0.22}_{0.22}$ & $0.07 \pm 0.01$ & $4.60 \pm 1.00$ & 25 \tnotex{tn:7} & ${6.4}$ & ${}^{+}_{-}$ & ${}^{0.2}_{0.2}$ & ${8.4}$ & ${}^{+}_{-}$ & ${}^{0.2}_{0.3}$ & ${18.2}$ & ${}^{+}_{-}$ & ${}^{10.8}_{10.7}$ & ${2.2}$ & ${}^{+}_{-}$ & ${}^{2.2}_{1.9}$ & ${8.7}$ & ${}^{+}_{-}$ & ${}^{0.7}_{1.6}$ & $0.0$ & ${18.2}$ & ${}^{+}_{-}$ & ${}^{52.9}_{5.3}$ & ${1.6}$ & ${}^{+}_{-}$ & ${}^{11.6}_{0.4}$ & ${8.5}$ & ${}^{+}_{-}$ & ${}^{2.1}_{0.4}$ & $0.8$ \\
Ursa Major II & ${3.62}$ & ${}^{+}_{-}$ & ${}^{0.31}_{0.34}$ & $0.15 \pm 0.02$ & $6.70 \pm 1.40$ & 20 \tnotex{tn:7} & ${7.0}$ & ${}^{+}_{-}$ & ${}^{0.2}_{0.2}$ & ${8.1}$ & ${}^{+}_{-}$ & ${}^{0.2}_{0.2}$ & ${23.7}$ & ${}^{+}_{-}$ & ${}^{8.4}_{12.4}$ & ${3.3}$ & ${}^{+}_{-}$ & ${}^{1.7}_{2.6}$ & ${9.1}$ & ${}^{+}_{-}$ & ${}^{0.4}_{1.3}$ & $0.0$ & ${65.0}$ & ${}^{+}_{-}$ & ${}^{8.6}_{48.4}$ & ${10.7}$ & ${}^{+}_{-}$ & ${}^{3.7}_{9.1}$ & ${10.4}$ & ${}^{+}_{-}$ & ${}^{0.2}_{2.0}$ & $0.6$ \\
Coma Berenices & ${3.58}$ & ${}^{+}_{-}$ & ${}^{0.27}_{0.29}$ & $0.08 \pm 0.01$ & $4.60 \pm 0.80$ & 59 \tnotex{tn:7} & ${6.4}$ & ${}^{+}_{-}$ & ${}^{0.2}_{0.2}$ & ${8.3}$ & ${}^{+}_{-}$ & ${}^{0.2}_{0.2}$ & ${17.2}$ & ${}^{+}_{-}$ & ${}^{9.6}_{8.5}$ & ${2.0}$ & ${}^{+}_{-}$ & ${}^{2.0}_{1.6}$ & ${8.6}$ & ${}^{+}_{-}$ & ${}^{0.7}_{1.3}$ & $0.0$ & ${18.2}$ & ${}^{+}_{-}$ & ${}^{49.2}_{0.0}$ & ${1.6}$ & ${}^{+}_{-}$ & ${}^{9.7}_{0.0}$ & ${8.5}$ & ${}^{+}_{-}$ & ${}^{2.0}_{0.0}$ & $1.1$ \\
Bootes II & ${3.02}$ & ${}^{+}_{-}$ & ${}^{0.38}_{0.38}$ & $0.05 \pm 0.02$ & $10.50 \pm 7.40$ & 5 \tnotex{tn:13} & ${6.9}$ & ${}^{+}_{-}$ & ${}^{0.5}_{0.7}$ & ${9.5}$ & ${}^{+}_{-}$ & ${}^{0.6}_{0.7}$ & ${63.0}$ & ${}^{+}_{-}$ & ${}^{7.1}_{61.7}$ & ${8.9}$ & ${}^{+}_{-}$ & ${}^{5.3}_{8.8}$ & ${10.4}$ & ${}^{+}_{-}$ & ${}^{0.3}_{4.9}$ & $0.1$ & ${18.2}$ & ${}^{+}_{-}$ & ${}^{55.4}_{15.7}$ & ${1.6}$ & ${}^{+}_{-}$ & ${}^{12.8}_{0.9}$ & ${8.5}$ & ${}^{+}_{-}$ & ${}^{2.2}_{2.0}$ & $0.3$ \\
Willman 1 & ${3.02}$ & ${}^{+}_{-}$ & ${}^{0.42}_{0.48}$ & $0.03 \pm 0.01$ & $4.30 \pm 1.80$ & 15 \tnotex{tn:14} & ${5.8}$ & ${}^{+}_{-}$ & ${}^{0.3}_{0.5}$ & ${9.2}$ & ${}^{+}_{-}$ & ${}^{0.4}_{0.5}$ & ${58.8}$ & ${}^{+}_{-}$ & ${}^{11.3}_{54.8}$ & ${11.3}$ & ${}^{+}_{-}$ & ${}^{3.0}_{11.1}$ & ${10.4}$ & ${}^{+}_{-}$ & ${}^{0.3}_{4.2}$ & $0.0$ & ${18.2}$ & ${}^{+}_{-}$ & ${}^{55.4}_{15.7}$ & ${1.6}$ & ${}^{+}_{-}$ & ${}^{12.8}_{0.9}$ & ${8.5}$ & ${}^{+}_{-}$ & ${}^{2.2}_{2.0}$ & $0.9$ \\
Segue II & ${2.94}$ & ${}^{+}_{-}$ & ${}^{0.16}_{0.16}$ & $0.03 \pm 0.00$ & $3.40 \pm 1.85$ & 5 \tnotex{tn:15} & ${5.8}$ & ${}^{+}_{-}$ & ${}^{0.4}_{0.6}$ & ${8.8}$ & ${}^{+}_{-}$ & ${}^{0.4}_{0.6}$ & ${22.0}$ & ${}^{+}_{-}$ & ${}^{36.9}_{20.6}$ & ${3.1}$ & ${}^{+}_{-}$ & ${}^{8.2}_{3.0}$ & ${9.0}$ & ${}^{+}_{-}$ & ${}^{1.4}_{3.6}$ & $0.0$ & ${18.2}$ & ${}^{+}_{-}$ & ${}^{55.4}_{15.7}$ & ${1.6}$ & ${}^{+}_{-}$ & ${}^{12.8}_{0.9}$ & ${8.5}$ & ${}^{+}_{-}$ & ${}^{2.2}_{2.0}$ & $0.3$ \\
Segue (I) & ${2.54}$ & ${}^{+}_{-}$ & ${}^{0.39}_{0.41}$ & $0.03 \pm 0.01$ & $3.90 \pm 0.80$ & 24 \tnotex{tn:16} & ${5.8}$ & ${}^{+}_{-}$ & ${}^{0.2}_{0.2}$ & ${9.0}$ & ${}^{+}_{-}$ & ${}^{0.3}_{0.3}$ & ${37.5}$ & ${}^{+}_{-}$ & ${}^{21.3}_{28.8}$ & ${6.2}$ & ${}^{+}_{-}$ & ${}^{5.0}_{5.9}$ & ${9.8}$ & ${}^{+}_{-}$ & ${}^{0.6}_{2.5}$ & $0.0$ & ${18.2}$ & ${}^{+}_{-}$ & ${}^{49.2}_{7.7}$ & ${1.6}$ & ${}^{+}_{-}$ & ${}^{9.7}_{0.5}$ & ${8.5}$ & ${}^{+}_{-}$ & ${}^{2.0}_{0.7}$ & $3.3$ \\

\bottomrule
\end{tabular}}
}

\begin{tablenotes}
 \item[1] \label{tn:1} \citet{Ibata1997}
 \item[2] \label{tn:2} \citet{Walker2009Fornax}
 \item[3] \label{tn:3} \citet{Mateo2008Leo}
 \item[4] \label{tn:4} \citet{Koch2007}
 \item[5] \label{tn:5} Spencer et al. (in prep.)
 \item[6] \label{tn:6} \citet{Walker2007bLetter}
 \item[7] \label{tn:7} \citet{Simon2007}
 \item[8] \label{tn:8} \citet{Torrealba2016}
 \item[9] \label{tn:9} \citet{CaldwellWalker2017}
 \item[10] \label{tn:10} \citet{Aden2009Hercules}
 \item[11] \label{tn:11} \citet{Koposov2011}
 \item[12] \label{tn:12} \citet{WalkerBelokurov2009}
 \item[13] \label{tn:13} \citet{Koch2009}
 \item[14] \label{tn:14} \citet{Martin2007}
 \item[15] \label{tn:15} \citet{Belokurov2009}
 \item[16] \label{tn:16} \citet{Geha2009}
\end{tablenotes}

\end{threeparttable}
\end{table}
\end{landscape}
}

\subsection{The tight correlation of enclosed mass and luminosity}
\label{sec:LGhalflightmass}
\begin{figure} 	 
  \centering
\includegraphics[width=\columnwidth]{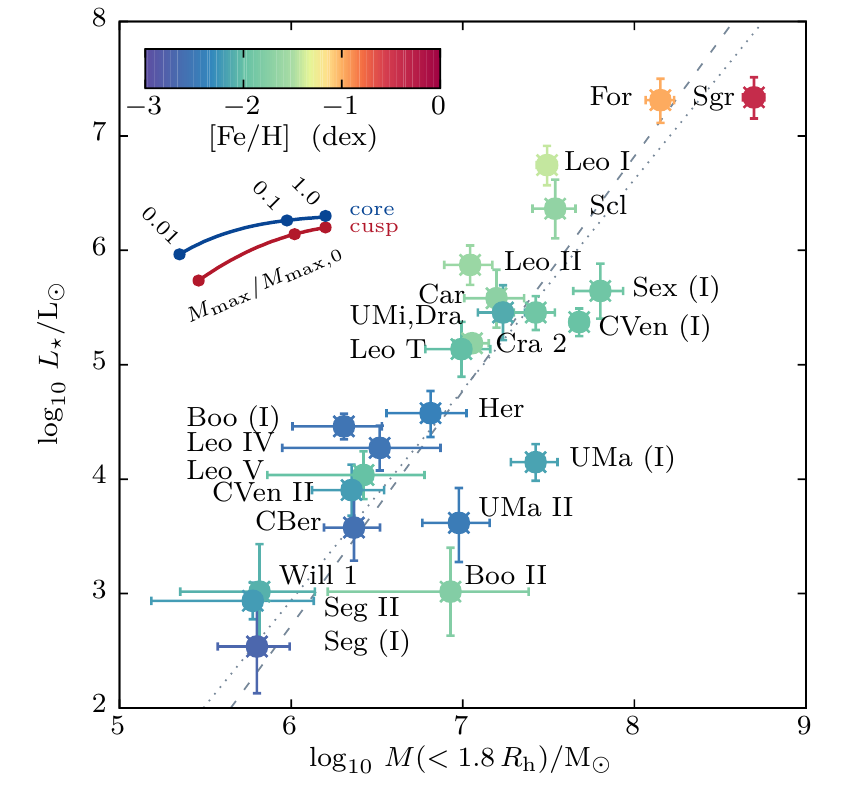}
\caption{Luminosity $L_\star$ as a function of the mass $M(<1.8\,\Rh)$ enclosed within 1.8 half-light radii for Milky Way dwarf galaxies.
Masses are estimated using the minimum-variance estimator (equation~\ref{eq:minvariance}). The colour-coding shows metallicity $\mathrm{Fe/H}$. Uncertainties are estimated by Monte Carlo sampling of symmetric Gaussian errors on $\langle \sigma^2_\mathrm{los} \rangle$ and $\Rh$. The blue (red) solid curve shows the evolution along tidal tracks (see Appendix \ref{appendix:tidal}) for a stellar population with $R_\mathrm{h,0}/r_\mathrm{max,0} = 1/10$ embedded in a cored (cuspy) DM profile. {The dashed (dotted) grey lines are power-law fits excluding (including) the Sagittarius dSph, see text}. Numerical values are listed in table~\ref{tab:results}.}
\label{fig:Menclosed}
\end{figure}

In this section we show that enclosed mass $M(<1.8\, \Rh)$ and luminosity $L_\star$ of Milky Way dSph galaxies are tightly correlated, and that the effect of the tidal field of the Milky Way is small for the evolution of this correlation since accretion of the dwarfs. To estimate the enclosed mass within $\bar \lambda = 1.8$ half-light radii $\Rh$, we use the minimum-variance $\mu=$const estimator (equation~\ref{eq:minvariance}). Fig.~ \ref{fig:Menclosed} shows the tight correlation between $L_\star$ and $M(<1.8\, \Rh)$, and numerical values are listed in Table~\ref{tab:results}. 
The faintest dwarfs of the sample with $L_\star \sim \unit[10^3]{L_{\odot}}$ have enclosed masses of $\sim \unit[10^6]{M_{\odot}}$, whereas the Fornax and Sagittarius\footnote{The Sagittarius dwarf galaxy is close to pericentre and perturbed by tidal forces of the Milky Way halo and disc \citep[see e.g.][]{Penarrubia2009,Lokas2010Sagittarius}. Our mass estimates are derived under the assumption of virial equilibrium of the stellar tracer component, a requirement which is not guaranteed to hold for Sagittarius.} dwarf galaxies with $L_\star \sim \unit[10^7]{L_{\odot}}$ have enclosed masses of the order $\sim \unit[10^{8}]{M_{\odot}}$.
The mean mass-to-light ratios $\Upsilon_\star$ averaged within the respective half-light radii therefore decrease with luminosity and span a range of $1 \lesssim \declog \Upsilon_\star \lesssim 3$. {This correlation has already been noted by \citet[][fig. 9]{Amorisco2011}, using the mass $M(<1.7\, \Rh)$ enclosed within $1.7$ half-light radii, and \citet{McConnachie2012}, applying the \citet{Walker2009} mass estimator and plotting masses enclosed within the half-light radius $M(<\Rh)$ as a function of absolute $V$-band magnitude.}
Note that the luminosity $L_\star$ is not used in the inference of $M(<1.8\, \Rh)$, distinguishing this correlation from the relation between enclosed mass $M(<\Rh)$ and half-light radius $\Rh$ discussed in \citet{Walker2009}.
{The correlation between $L_\star$ and $M(<1.8\, \Rh)$ can be fit by a power law of the form $L_\star/\unit[]{L_\odot} = 10^a \times \left[M(<1.8\, \Rh)/\unit[]{M_\odot}\right]^b$. Assuming symmetric uncertainties on $\declog\,M(<1.8\, \Rh)$ and $\declog\,L_\star$, we find \mbox{$a=-9.6\pm2.3$}, \mbox{$b=2.1\pm0.3$} when excluding the Sagittarius dSph from the fit, and \mbox{$a=-8.1\pm1.7$}, \mbox{$b=1.8\pm0.2$} when including it.}
The colour-coding of Fig.~\ref{fig:Menclosed} shows the metallicity Fe/H of the dwarfs, which also increases with $M(<1.8\, \Rh)$. 

Our results suggest that the correlation of $M(<1.8\, \Rh)$ and $L_\star$ is mainly driven by internal processes, i.e. star formation and feedback, and that tides through by Milky Way halo do have little effect on the relation. Solid lines in Fig.~\ref{fig:Menclosed} show the evolution of luminosity and enclosed mass along tidal evolutionary tracks (see Appendix \ref{appendix:tidal}), i.e. as a function of the remnant mass fraction $M_\mathrm{max}/M_\mathrm{max,0}$, separately for the assumption of stellar populations embedded in cuspy and cored DM haloes. The initial segregation of the stellar populations is chosen to be $R_\mathrm{h,0}/r_\mathrm{max,0} = 1/10$. Dwarf galaxies evolving along these tidal tracks can lose up to two orders of magnitude of their mass $M_\mathrm{max}$ enclosed within $\rmax$, or equivalently of their total mass under the assumption that the profile shape doesn't change, but experience a decrease in $M(<1.8\,\Rh)$ by only one order of magnitude. This is related to $\rmax$ decreasing during tidal stripping (see top panel of Fig.~\ref{fig:tidaltracks}), which causes the mass enclosed within the central regions of the halo to decrease less rapidly than the total mass. As a consequence, the relation of luminosity and enclosed mass of Fig.~\ref{fig:Menclosed} is not expected to broaden significantly during the tidal evolution of the dwarfs after accretion, and this holds both under the assumption of cuspy and cored DM halo profiles. However, internal processes like star formation will likely have an effect on the relation also after accretion, with star formation shown to be ongoing in some dSphs also after accretion - see e.g. \citet{deBoer2012Fornax}, discussing that the latest star formation in the Fornax dwarf happened as recently as \unit[250]{Myrs} ago. 

\begin{figure*}
  \centering
\includegraphics[width=\textwidth]{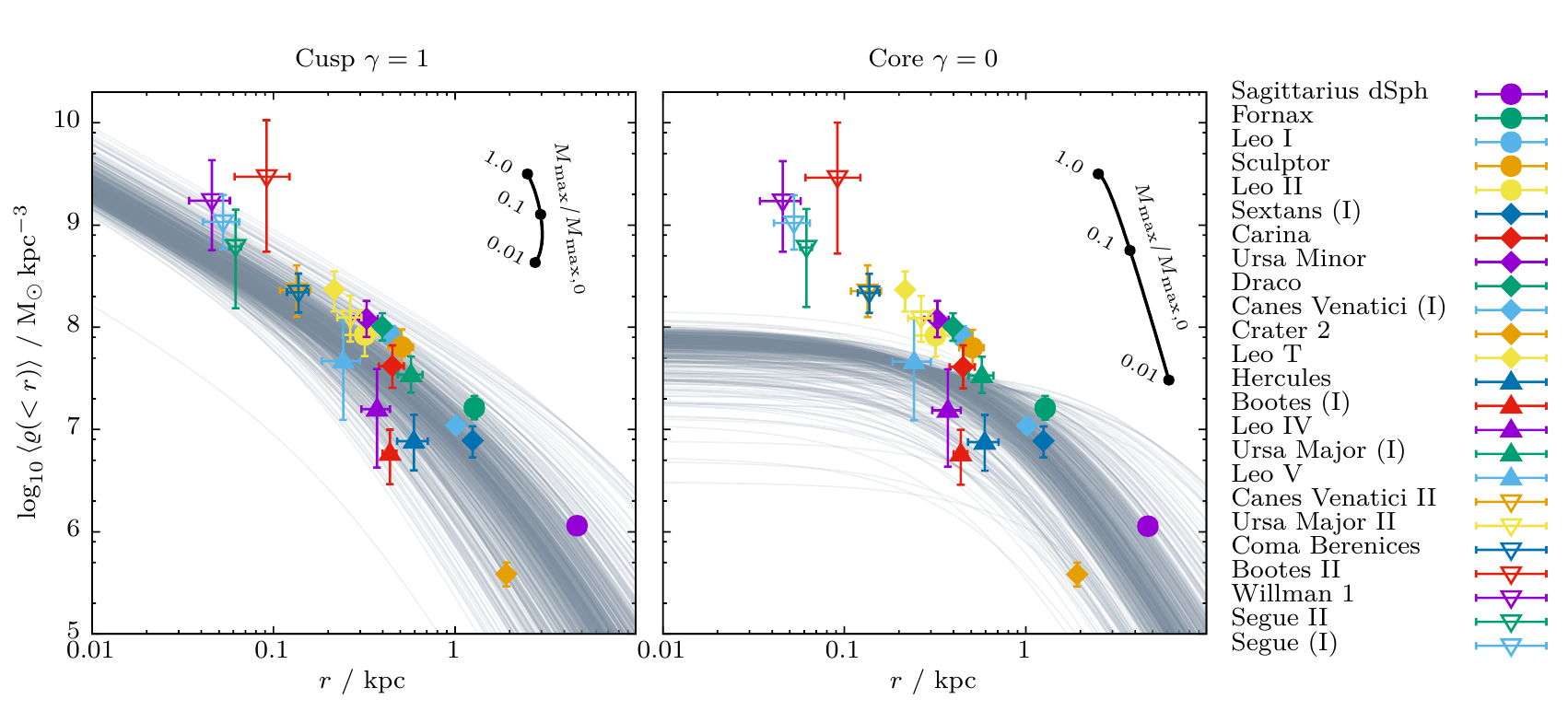}
\caption{Mean densities $\varrho(<1.8\, \Rh)$ within $1.8$ half-light radii $\Rh$ of Milky Way dSphs, estimated using the minimum-variance mass estimator (equation~\ref{eq:minvariance}). Grey curves show the mean density profiles of cuspy ($\gamma=1$, left-hand panel) and cored ($\gamma=0$, right-hand panel) haloes of the controlled cosmological simulations introduced in section \ref{sec:methodcompenclosed}. These curves assume $\{\alpha,\beta,\gamma\}=\{1,5,\gamma\}$ profiles with $\{\rmax,\vmax\}$ as fitted to the simulated haloes. Note that the ultrafaint Milky Way dwarfs are too dense to be compatible with any of the simulated cored DM haloes.}
\label{fig:densradius}
\end{figure*}
\begin{figure*} 	 
  \centering
\includegraphics[width=\textwidth]{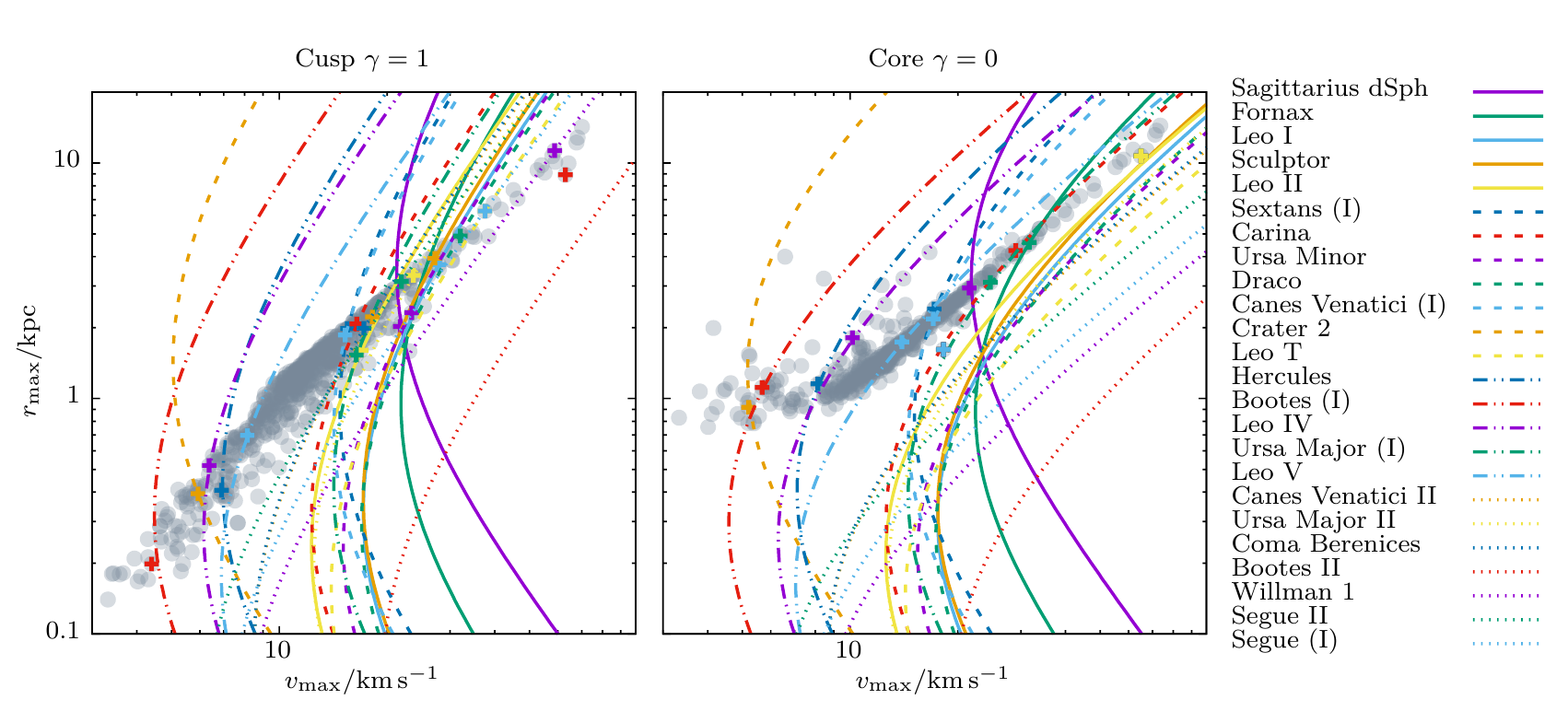}
\caption{$\{\rmax,\vmax\}$ degeneracy curves of the DM haloes of Milky Way dSphs obtained by assuming cuspy (left-hand panel, $\gamma=1$) and cored (right-hand panel, $\gamma=0$) DM density profiles for tidally stripped systems ($\alpha=1$, $\beta=5$ in equation~\ref{eq:betagammaprofile}). $\{\rmax,\vmax\}$ values of subhaloes of our controlled cosmological simulations are shown by grey filled circles. Crosses indicate the minimum-$\chi^2$ estimates (see equation~\ref{equ:medianChi2sigma}). Note that the $\{\rmax,\vmax\}$ curves of the ultrafaint Milky Way dwarfs result incompatible with any of the simulated cored DM haloes. }
\label{fig:LGvsCCCP2}
\end{figure*}

\subsection{Confronting measured densities and simulated haloes}
When comparing the mean densities of ultrafaint Milky Way dwarfs to those of cored DM haloes taken from our re-simulations of the Aquarius A2 merger tree, we find that none of the simulated cored haloes are dense enough to host an ultrafaint Milky Way dwarf. We compute the mean density, averaged within a spherical volume of radius $1.8\Rh$, using the minimum variance mass estimator (equation \ref{eq:minvariance}):
\begin{equation}
 \langle  \varrho(<1.8\,\Rh) \rangle = {M(<1.8\, \Rh)} (4\pi/3)^{-1} (1.8\,\Rh)^{-3} .
\end{equation}
Fig.~\ref{fig:densradius} shows the mean densities $\varrho(<1.8\,\Rh) \rangle$ as a function of half-light radius $\Rh$, as well as the mean density profiles of the cuspy ($\gamma=1$) and cored ($\gamma=0$) DM haloes of the controlled cosmological simulations of the formation of Milky Way-like haloes described in section \ref{sec:methodcompenclosed}. These profiles are obtained by determining $\{\rmax,\vmax\}$ for each simulated DM halo at $z=0$, and assuming an $\{\alpha,\beta,\gamma\}=\{1,5,\gamma\}$ profile with those values of $\{\rmax,\vmax\}$. 
Whereas for the re-simulation using cuspy subhaloes there are subhaloes with densities as high as the measured densities $\varrho(<1.8\,\Rh) \rangle$ of Milky Way dwarfs, in the case of cored simulations, none of the simulated haloes is dense enough to be compatible with the ultrafaint galaxies in the sample (see right-hand panel of Fig.~\ref{fig:densradius}). Also shown in Fig.~\ref{fig:densradius} is the evolution along tidal tracks (see Appendix \ref{appendix:tidal}) for a stellar population with an initial segregation of $R_\mathrm{h,0}/r_\mathrm{max,0} = 1/10$, assuming $L_\star \propto M_\star$. With half-light radii staying near constant, the evolution of the mean density for cuspy systems reflects the previously discussed behaviour that a system being stripped two orders of magnitude of its total mass only loses one order of magnitude of its enclosed mass $M(<1.8 \Rh)$, and therefore also $\langle\varrho(<1.8\,\Rh) \rangle$ drops by only one order of magnitude. The situation is notably different for dSphs embedded in cored DM haloes, which have half-light radii that expand during tidal stripping (see bottom panel of Fig.~\ref{fig:tidaltracks}). For cored systems, losing two orders of magnitude of their total mass results in the average central density $\langle  \varrho(<1.8\,\Rh) \rangle$ to drop by two orders of magnitude as well. As a consequence, if the ultrafaint galaxies of the sample did lose mass through tides previously, they must have been even denser in the past, rendering them even more incompatible with the cored DM haloes of our simulations\footnote{It is interesting to note that increasing the mass of the host halo will not ease the difference between the measured densities and those of cored simulated haloes: whereas for cuspy density profiles, those systems with the highest total mass also have the largest densities, this does not hold for cored haloes. For the case of tidally evolved, cored $\{\alpha,\beta,\gamma\}=\{1,5,0\}$ profiles, from equation~\ref{eq:cumulativemassbetagamma} we find for the central density $\varrho(0) = \varrho_s \propto M_\mathrm{tot}/a^3$. An empirical power-law fit to the relation between $M_\mathrm{tot}$ and the scale radius $a$, derived from the Aquarius A2 merger tree for systems with mass $\geq \unit[10^8]{M_{\odot}}$, gives $a \propto M_\mathrm{tot}^{0.4}$. Consequently for these mass scales, the central density decreases with increasing mass. {This result holds for the mass-size (or equivalently $\{\rmax,\vmax\}$) relation of the Aquarius merger tree. In principle DM models which result in cored haloes may have different clustering properties and $\{\rmax,\vmax\}$ relations. Choosing a different mass-size relation for cored haloes at infall will effect the densities of the simulated haloes.}}. This tension is a consequence of our choice of core size of the simulated haloes. The cored subhaloes injected in the evolving host potential are modelled as \citet{dehnen1993} density profiles with $\gamma=0$, i.e. density profiles where the core size is equal to the scale radius, chosen to highlight the different tidal evolution of cored substructures compared to cuspy ones. Our findings therefore suggest that the ultra faint dwarfs require core sizes that are much smaller than the dark matter scale radius.

\subsection{(Total) halo mass - stellar mass relation}
\label{sec:dwarf}
Extrapolating total halo masses from measured enclosed masses is an uncertain endeavour: whereas enclosed masses within multiples of the half-light radius are well constrained by measurements of $\langle \sigma^2_\mathrm{los}\rangle$ and $\Rh$, constraints on the total halo mass are very weak ($\lambda \gg 1$ in Fig.~\ref{fig:lambdaBias}). Total halo masses are therefore commonly extrapolated from enclosed masses under the assumption of median cosmological mass - concentration relations obtained for cuspy \citet{nfw1997} profiles (see e.g. the discussion in \citet{Strigari2008}, making use of the \citet{Bullock2001MvirVmax} relation between virial mass $M_\mathrm{vir}$ and $\vmax$ derived using cosmological $N$-body simulations). Tidally stripped systems do have density profiles with steep outer slopes \citep[$\beta \approx 5$, see][]{Penarrubia2009,Penarrubia2010} and convergent total masses, in contrast to the \citet{nfw1997} profile of field haloes with outer slope $\beta=3$ and a divergent total mass. Adding to this, following the tidal evolution of low-mass subhaloes is a challenging task for traditional cosmological simulation with fixed $N$-body particle mass, and recent studies call for caution regarding numerical convergence of these simulations \citep{vdb2018}.

We infer the total halo mass of Milky Way dwarf galaxies by fitting the observed line-of-sight velocity dispersion $\langle \sigma^2_\mathrm{los} \rangle$ to $N$-body DM subhaloes extracted from a re-simulation of the Aquarius A2 merger tree, as described in section \ref{sec:totalmass}. 
Fig.~\ref{fig:LGvsCCCP2} shows $\{\rmax,\vmax\}$ degeneracy curves constrained by the measurements of velocity dispersion $\langle \sigma^2_\mathrm{los} \rangle$ and half-light radius $\Rh$ for Milky Way dSph galaxies assuming cuspy ($\gamma=1$) or cored ($\gamma=0$) DM density profiles for tidally stripped systems ($\alpha=1$, $\beta=5$ in equation~\ref{eq:betagammaprofile}). The $\{\rmax,\vmax\}$ values of $N$-body subhaloes at $z=0$ of our controlled simulations are depicted by grey filled circles, and the haloes corresponding to the minimum-$\chi^2$ estimates (see equation~\ref{equ:medianChi2sigma}) are indicated using crosses.
Table~\ref{tab:results} lists the inferred masses and structural parameters for the minimum-$\chi^2$ estimate.
For the case of cored DM profiles, the dwarfs {Leo I} and Segue (I) have $\chi^2_\mathrm{min} > 2$ and therefore result incompatible with all simulated subhaloes. Note that for cored systems, the $\{\rmax,\vmax\}$ degeneracy curves corresponding to ultrafaint galaxies do not intersect with any of the simulated cored DM haloes. Nevertheless the large observational uncertainties of the kinematics of these ultrafaint systems, for the case of cored DM profiles, we have $\chi^2_\mathrm{min} \gtrsim 0.3$ for all dwarfs with $L_\star < \unit[10^4]{L_{\odot}}$.

Based on the inferred total halo masses $M_\mathrm{tot}$, we study the stellar- to halo mass relation of Milky Way satellite galaxies at $z=0$.
Using the rough approximation that on average $M_\star/\mathrm{M_{\odot}} \approx 1.5\,L_\star/\mathrm{L_{\odot}}$ \citep[e.g.][]{Martin2008MWsat}, Fig. \ref{fig:abundancematch} shows the stellar- and total halo masses of MW dwarfs. {While this approximation for $M_\star/\mathrm{M_{\odot}}$ is uncertain by a factor of the order of unity, the general trends discussed below regarding Fig. \ref{fig:abundancematch}, spanning over several orders of magnitude in luminosity and mass, will not be effected by choosing a different mean stellar mass- luminosity ratio.}
Halo masses inferred under the assumption of cuspy and cored subhaloes are shown in the top and bottom panels, respectively. 
Errorbars show the minimum and maximum masses of simulated subhaloes which satisfy $\chi^2 < \chi^2_\mathrm{min} +1$. The stellar- to halo mass relations by \citet{Moster2010}, \citet{Behroozi2013} and \citet{SawalaFrenk2016} are plotted using dashed grey lines, showing $M_{200}$ as a proxy to the total halo mass (which for NFW, i.e. non-tidally stripped profiles with $\beta=3$, is diverging). 
For systems with $\declog L_\star/\mathrm{L_{\odot}} \gtrsim 5$, our findings are compatible with all three stellar- to halo mass relations shown. For the case of the Sagittarius and the Fornax dSphs, our inferred masses are smaller by one order of magnitude than the halo mass predictions for field haloes given the observed luminosities. In general, for the case of cuspy subhaloes, all classical Milky Way dSphs have inferred masses at the low-mass end of the abundance matching relations shown. This discrepancy is consistent with our understanding of the tidal evolution of dSph galaxies: when being accreted on to a larger galaxy, dSph galaxies initially lose predominantly DM and not stars, which are embedded deeply inside the galaxies' potential and are therefore less prone to stripping \citep{penarrubia2008}. The tidal evolutionary tracks shown in Fig.~\ref{fig:abundancematch} and discussed in Appendix \ref{appendix:tidal} show that dSphs with both cuspy and cored DM profiles can be stripped one order of magnitude of their dynamical mass due to tides while losing less than ten per cent of their stellar mass.
Note also that the correlation between luminosity and enclosed mass $M(<1.8 \, \Rh)$ (see Fig.~\ref{fig:Menclosed}) is notably tighter than the correlation found between stellar mass and total halo mass. This could be a consequence of the tidal evolution of the dwarfs: if at infall stellar mass and total halo mass were more tightly correlated, tides may have stripped significant fractions of DM but not of stars, as consistent with the tidal tracks for highly segregated stellar populations.

\subsection{Puzzling halo masses for ultrafaint dwarfs}
\begin{figure} 	 
  \centering
  \includegraphics[width=8.5cm]{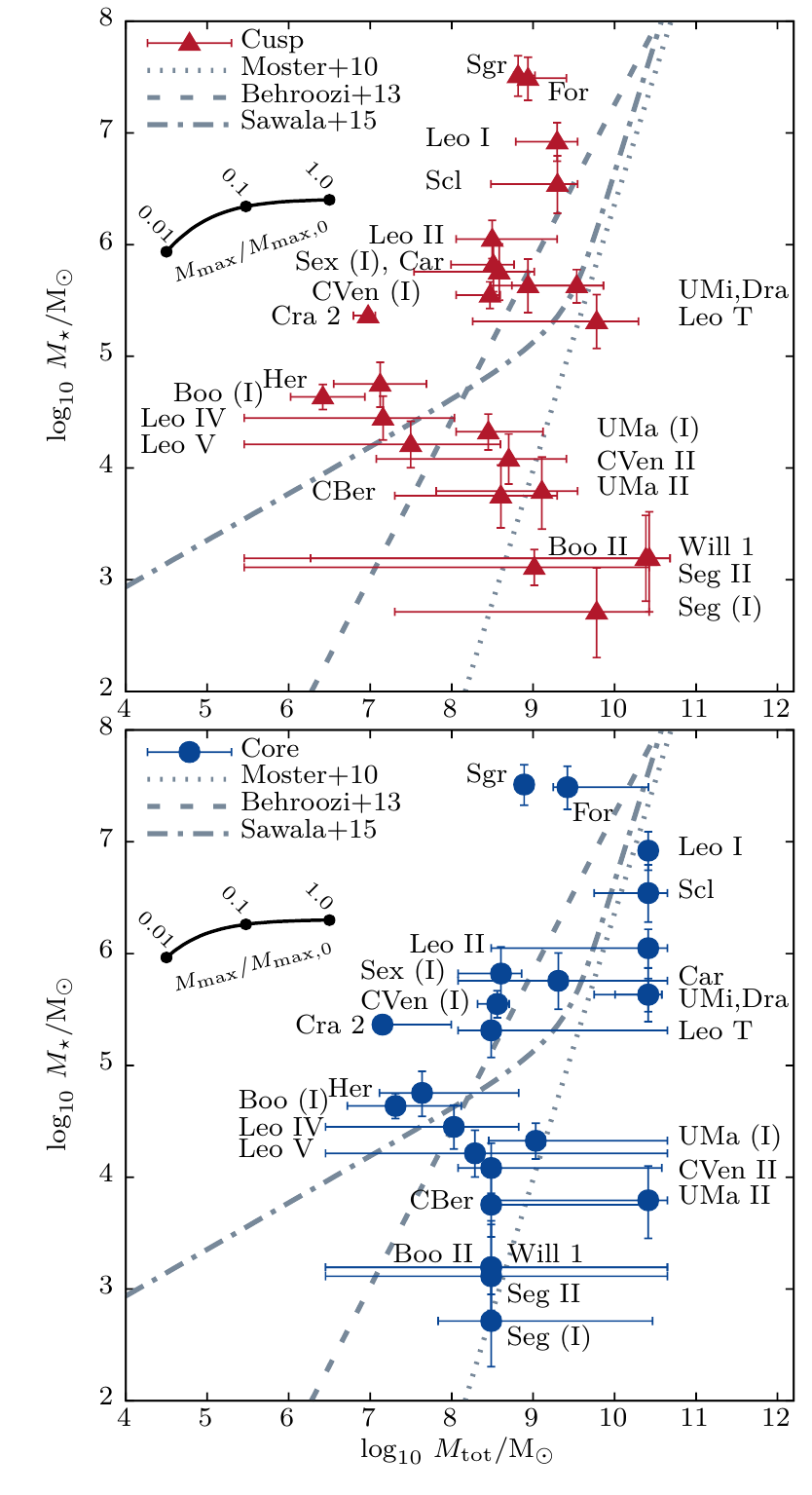} 
     \caption{Stellar- and total halo masses inferred for Milky Way dwarf galaxies \citep{McConnachie2012} by fitting the observed velocity dispersion $\langle \sigma_\mathrm{los}^2 \rangle$ to the catalogue of subhaloes of our re-simulations of the Aquarius A2 merger tree. 
     Abundance matching relations \citep{Moster2010, Behroozi2013, Sawala2015} are shown as grey dashed lines. The back solid lines show the median evolution of $M_\star$ following the tidal tracks of Appendix \ref{appendix:tidal} as a function of the fraction of remnant mass $M_\mathrm{max}$ enclosed within $\rmax$. }
  \label{fig:abundancematch}
\end{figure}

For systems with $\declog\, L_\star/\mathrm{L_{\odot}} \lesssim 5$, we find an anticorrelation between stellar mass and halo mass, associating the faintest galaxies to the most massive haloes. For cored systems, Milky Way dwarf galaxies with luminosities spanning between {$2 \lesssim \declog\, L_\star/\mathrm{L_{\odot}} \lesssim 4$} are associated to the DM haloes of mass of $\sim 10^{9} \mathrm{M_{\odot}}$, with an uncertainty of two orders of magnitude and $\chi^2_\mathrm{min} \gtrsim 0.3$. The corresponding $\{\rmax,\vmax\}$ degeneracy curves of Fig.~\ref{fig:LGvsCCCP2} do not intersect with any $\{\rmax,\vmax\}$ value measured from the cored simulations. 
Note that dwarf galaxies evolving along the tidal tracks shown in Fig.~\ref{fig:abundancematch} primarily lose DM and not stellar mass, and move away from the stellar mass- halo mass relation found for ultrafaints, i.e. tidal evolution cannot be at the origin of the observed anticorrelation.

There are however observational challenges which might be related to this counter-intuitive anticorrelation:
\begin{itemize}
\item Inflated masses might result from inflated velocity dispersion measurements: a striking example being the Hercules dSph, where \citet{Aden2009Hercules} found a reduction by roughly half from $7.33 \pm 1.08 \mathrm{km\,s^{-1}}$ to $3.72 \pm 0.91\, \mathrm{km\,s^{-1}}$ after removing foreground contaminants. {In comparison, \citet{Simon2007} computed a dispersion of $5.1 \pm 0.9\, \mathrm{km\,s^{-1}}$.}
\item Similarly, peculiar velocities due to binary motion add to the observed velocity dispersion \citep[see e.g.][]{McConnachieCote2010}.
\item  Low-number statistics for systems with few stellar tracers add to the uncertainty on the estimated dispersion. Table \ref{tab:results} lists the number $N$ of stars with measured velocities used to infer the dispersion $\sigma_\mathrm{los}$. Most dwarfs with luminosity $L_\star \lesssim \unit[10^5]{L_\odot}$ have fewer than 50 member stars with measured velocities, the only exceptions being Bootes (I) with $\sim$ 50 member stars and Coma Berenices with 59 stars. The ultrafaint dwarfs in the sample with  $L_\star \lesssim \unit[10^3]{L_\odot}$ have less than 25 member stars with measured velocities, Bootes II and Segue II having only five member stars each. Following \citet[][table 1]{Laporte2018}, for seven stars and a ratio between measured dispersion and measurement error of $\sigma/{\varepsilon_v} = 2.0$, the uncertainty on the inferred velocity dispersion due to sample size is $\sim 50$ per cent, whereas for 51 member stats, the uncertainty is $\sim$ 25 per cent.
\end{itemize}

Furthermore, there are several systematic aspects of our method which call for caution.
\begin{itemize}
\item All our analysis is based on the virial theorem and thereby assumes dynamical equilibrium of the underlying stellar tracer component. Some dSphs do show extra-tidal (i.e. possibly non-equilibrium) features, and the applicability of the virial theorem is then limited to the virialized core of the systems.
\item We infer total halo masses of Milky Way dwarfs by selecting subhaloes extracted from simulations of Milky Way-like DM haloes compatible with the observed kinematics. For this purpose, we re-simulated a single merger tree of the Aquarius simulation. Cosmic variance will allow for a wider range of accretion scenarios, and possibly for a wider range of $\{\rmax,\vmax\}$ values at $z=0$. This may increase the number of haloes compatible with the observed kinematics. {Moreover, we use the same $\{\rmax,\vmax\}$ relation  at infall for both cuspy and cored models, whereas in principle DM models that result in cored haloes may have different clustering properties and $\{\rmax,\vmax\}$ relations compared to their cuspy counterparts. }
\item The Aquarius A2 main halo has a virial mass of $\unit[1.84 \times 10^{12}]{M_{\odot}}$, which is larger by a factor $\sim 2$ than estimates of the (total) Milky Way mass derived by use of the timing argument \citep{Penarrubia2016Timing}. Therefore also the simulated subhaloes, which we use to break the $\{\rmax,\vmax\}$ degeneracy of Milky Way dSphs, might have on average higher masses than Milky Way subhaloes.
\item We infer masses separately under the assumption of cuspy and cored DM profiles, using a single choice of core size. Different core sizes are motivated by various physical models for core formation. Our choice to model the subhaloes at infall as \citet{dehnen1993} profiles with $\gamma=0$ and $\gamma=1$ are extreme cases, but as shown in Fig.~\ref{fig:abundancematch}, the impact of the assumed core size on the inferred total halo masses is relatively small. 
\end{itemize}

In this contribution, we made use of measured velocity dispersions and half-light radii of Milky Way dSphs to infer $\{\rmax,\vmax\}$ degeneracy curves, and used structural parameters $\{\rmax,\vmax\}$ of simulated DM subhaloes to break this degeneracy. This approach neglects orbital constraints available for Milky Way dSphs, and associates simulated DM haloes to Milky Way dSphs independently of their orbit or accretion history. 
In a follow-up contribution we will explore statistical methods to incorporate information on the distance measurements, radial velocities and proper motions into our fits in order to constrain the accretion history of Milky Way dSphs.
Instead of associating Milky Way dSphs to simulated DM subhaloes of a single merger tree, we will use a probabilistic approach to model the distribution of structural and orbital properties of DM subhaloes in Milky Way-like haloes, as introduced in \citet{Penarrubia2018}. 

{
\subsection{Comparison with other studies}
The masses enclosed within the luminous radii of dwarf galaxies have been extensively discussed in the literature, and the enclosed masses $M(<1.8\,\Rh)$ estimated using the minimum variance estimator (equation \ref{eq:minvariance}) listed in Table \ref{tab:results} differ at most by factors of order of unity from the mass estimates in \citet{Walker2009},  \citet{Wolf2010}, \citet{Amorisco2011}, \citet{McConnachie2012} when extrapolated to the same radius. The differences are due to i) differently motivated choices of the factors $\mu,\lambda$ (see equation \ref{eq:mu-general-form}) and ii) updated velocity dispersion measurements. 

Following \citet{Mateo1993}, \citet{Strigari2008} and \citet{Wolf2010} discuss a single mass scale of $\sim\unit[10^9]{M_{\odot}}$ for the total masses of all Milky Way dwarf galaxies, making use of cosmological mass-size relations for cuspy haloes to extrapolate the total mass from masses enclosed within \unit[300]{pc} and the deprojected half-light radius, respectively. This mass scale is in agreement with the total masses we derive for most Milky Way dSphs, however on the low-mass end, we derive total masses of $\sim\unit[10^7]{M_{\odot}}$ for Crater 2, Hercules and Leo IV and $\sim\unit[10^6]{M_{\odot}}$ for Bootes (I). In the following we compare our mass estimates (as listed in Table \ref{tab:results}) for four selected Milky Way dwarf galaxies against other studies.

\subsubsection{Fornax   $(L_\star \approx \unit[2 \times 10^7]{L_\odot})$}
We infer an enclosed mass of $M_\mathrm{est}(<1.8\,\Rh) = \unit[(1.4^{+0.3}_{-0.3})\times10^8]{M_\odot}$ for Fornax and, assuming a cuspy $\{\alpha,\beta,\gamma\}=\{1,5,1\}$ DM profile for tidally stripped systems, a total halo mass of $M_\mathrm{tot} = \unit[(8.7^{+16.9}_{-3.0})\times10^8]{M_\odot}$. For a cored $\{1,5,1\}$ DM profile, we find $M_\mathrm{tot} = \unit[(2.6^{+23.1}_{-0.9})\times10^9]{M_\odot}$.
For models where mass follows light, \citet{Lokas2009} obtain through Jeans modelling a total mass of $\unit[(1.6 \pm 0.1)\times10^8]{M_\odot}$, lower than our total mass estimate by an order of magnitude, and more similar to our estimated enclosed mass.
The best-fitting model of \citet{Diakogiannis2017} is as well a model where mass follows light, with total mass and uncertainty equal to the value of \citet{Lokas2009}. From a distribution function based approach \citet{Pascale2018} find total masses of $\unit[(2.1 \pm 0.1)\times10^8]{M_\odot}$ for models where mass follows light, $\unit[(5.8^{+1.9}_{-1.1})  \times10^9]{M_\odot}$ for a cuspy DM profile and  $\unit[(9.5^{+12}_{-2})  \times10^9]{M_\odot}$ for their best-fitting cored model. All DM halo models considered have convergent total masses, but outer slopes of $\beta = - \mathrm{d}\ln\varrho/\mathrm{d}\ln r \approx 3$. Consequently these systems have slowly converging total masses\footnote{Note that the NFW outer slope of $\beta\equiv3$ leads to a diverging total mass $M(<r/a) \propto \ln(r/a)$ for $r/a \gg 1$, with $a$ denoting the scale radius.}. We assume an outer slope of $\beta=5$ for tidally stripped systems, with much more rapidly converging total mass. The enclosed mass  $M(<1.7\,\Rh) = \unit[(1.4\pm0.1)\times10^8]{M_\odot}$ inferred by \citet{Pascale2018} is in good agreement with our estimate from the virial theorem using $\mu(\lambda=1.7)=3.46$ (see Fig.~\ref{fig:lambdaBias}): $M_\mathrm{est}(<1.7\,\Rh) = \unit[(1.3\pm0.3)\times10^8]{M_\odot}$.

\subsubsection{Sculptor   $(L_\star \approx \unit[2 \times 10^6]{L_\odot})$}
The enclosed mass we infer for Sculptor, $M(<1.8\,\Rh = \unit[(3.5^{+1.1}_{-0.9})\times10^7]{M_\odot}$, is again close to the total mass quoted by \citet{Lokas2009} for models where mass follows light, $\unit[(3.1 \pm 0.2)\times10^7]{M_\odot}$ . 
\citet{Strigari2017} use a distribution function based approach to conclude that observed stellar velocities in the Sculptor dwarf are consistent with an NFW DM halo with a maximum circular velocity of  $20 < \vmax/\kms < 35$.
This is similar to the range of maximum circular velocities we find under the assumption of cuspy $\{1,5,1\}$ profiles, $18 \lesssim \vmax/\kms \lesssim 32$. This corresponds to a total halo mass of $M_\mathrm{tot} =  \unit[(2.0^{+1.6}_{-1.7})\times10^9]{M_\odot}$. For cored $\{1,5,0\}$ profiles, we find $40 \lesssim \vmax/\kms \lesssim 74$, and $M_\mathrm{tot} = \unit[(2.6^{+1.9}_{-2.1})\times10^{10}]{M_\odot}$.

\subsubsection{Carina   $(L_\star \approx \unit[4 \times 10^5]{L_\odot})$}
For the Carina dwarf, we obtain an enclosed mass of $M_\mathrm{est}(<1.8\,\Rh) = \unit[(1.6^{+0.7}_{-0.6})\times10^7]{M_\odot}$, similar to the total mass inferred by \citet{Lokas2009} $\unit[(2.3 \pm 0.2)\times10^7]{M_\odot}$ under the assumption that mass follows light. Non-equilibrium $N$-body models of the Carina dwarf have been studied by \citet{Ural2015Carina}. They model the DM distribution of Carina separately under the assumption of cuspy ($\gamma=1$) and cored ($\gamma=0$) \citet{dehnen1993} profiles, i.e. $\{\alpha,\beta,\gamma\} = \{1,4,\gamma\}$. For the cuspy model, they find a total halo mass at $z=0$ of $4.0_{-2.4}^{+4.1} \times \unit[10^8]{M_{\odot}}$. This result is consistent with the total mass we find under the assumption of virial equilibrium and a $\{\alpha,\beta,\gamma\} = \{1,5,1\}$ DM profile, namely $3.8_{-3.5}^{+6.4} \times \unit[10^8]{M_{\odot}}$.
For the cored model, \citet{Ural2015Carina} find a total mass of $3.5_{-2.3}^{+3.9} \times \unit[10^8]{M_{\odot}}$. 
Our cored model for the Carina dSph is particularly poorly constrained (see right-hand panel of Fig.~\ref{fig:LGvsCCCP2}), and a large range of simulated haloes with masses spanning \mbox{$1.2 \times 10^8~-~ 4.5 \times 10^{10}\,\mathrm{M_{\odot}}$} result compatible with the observed kinematics.

\subsubsection{Hercules   $(L_\star \approx \unit[4 \times 10^4]{L_\odot})$}
On the low-mass end of Milky Way satellites, for the Hercules dwarf, we find an enclosed mass of $M_\mathrm{est}(<1.8 \Rh)/\mathrm{M_{\odot}} = 6.5_{-2.9}^{+4.0} \times 10^6$. To compare this value against the mass $M<(\unit[0.3]{kpc})/\mathrm{M_{\odot}}=1.9^{+1.1}_{-0.8}\times10^6$ inferred by \citet{Aden2009Hercules}, we use the general form of the virial mass estimator (equation~\ref{eq:massEst}) with $\mu(\lambda=\unit[0.3]{kpc}/\Rh) \approx 2.7$ (see Fig.~\ref{fig:lambdaBias}). We find $M_\mathrm{est}(<\unit[0.3]{kpc})/\mathrm{M_{\odot}} = 2.6_{-1.2}^{+1.8} \times 10^6$, in good agreement with the findings of \citet{Aden2009Hercules}.

} 

\section{Conclusions and Summary}
\label{sec:conclusionsSummary}
Mass estimators for pressure-supported systems play an important role in constraining the distribution of dark matter (DM) on the scale of dSphs. Kinematic data for these systems is often limited to velocity dispersion measurements $\sigma_\mathrm{los}(r)$ along the line of sight. The challenge lies in constructing an estimator which does not rely on strong assumptions about quantities inaccessible to observation.
In this contribution, we construct an estimator for the mass enclosed within 1.8 multiples of the projected half-light radius $\Rh$ of the stellar tracer population (equation~\ref{eq:minvariance}):
\[
 M_\mathrm{est}(<1.8\,\Rh) \approx 3.5 \times 1.8 \, \Rh \, G^{-1} \, \langle \sigma_\mathrm{los}^2 \rangle ~, 
\]
where by $\langle \sigma^2_\mathrm{los}\rangle = 2K$ we denote the luminosity-averaged squared line-of-sight velocity dispersion of the stellar tracer population. 
This estimator is based on the projected virial theorem and minimizes the uncertainty on the inferred masses arising from our ignorance on (i) the central slope $\gamma$ of the DM profile, as well as on (ii) how deeply embedded the stellar tracer population is within the DM halo. The estimator has been tailored to give accurate masses for tidally stripped dwarf galaxies, which follow density profiles that scale as $\varrho(r) \propto r^{-5}$ at large radii \citep{Penarrubia2009, Penarrubia2010}

The use of the projected virial theorem has several advantages over the Jeans equations for the construction of mass estimators:
\begin{itemize}
 \item Our method does not suffer from the mass - anisotropy degeneracy. The Jeans equations depend on information about the anisotropy of the velocity dispersion, parametrized by the function $\beta(r) \equiv 1 - {\sigma^2_\mathrm{los}(r)}/{\sigma^2_{\bot}(r)}$, therefore requiring knowledge about the velocity dispersion component $\sigma_{\bot}(r)$ orthogonal to the line of sight. The form of $\beta(r)$ might be different for each stellar population, and could be more complicated than a simple monotonic function of radius. Our ignorance of $\beta(r)$ gives rise to the infamous mass - anisotropy degeneracy. The projected virial theorem \mbox{$2K_\mathrm{los}+W_\mathrm{los}=0$} instead makes use of the luminosity-averaged squared line-of-sight velocity dispersion $\langle \sigma^2_\mathrm{los}\rangle = 2K$, accessible to observation.
 \item Systematic biases of inferred enclosed masses $M_\mathrm{est}(<1.8\,\Rh)$ and derived central slopes $\gamma$ are straightforward to estimate: they follow directly from the assumptions on the DM and stellar density profiles, and do not rely on assumptions on the difficult to constrain form of $\beta(r)$.
 \item The average squared dispersion $\langle \sigma_\mathrm{los}^2 \rangle$ is a sum over all stars and does not require data to be binned. It therefore can be robustly computed also for systems with a low number of stars - carefully modelling the uncertainties due to sample size as pointed out by \citet{Laporte2018}. 
\end{itemize}
However, the use of the virial theorem requires to specify a (family of) DM mass distributions, i.e. the estimated enclosed masses are model dependent. The bias of the derived masses is predominantly driven by how deeply embedded the stellar tracer is within the DM halo. 
For this reason, masses enclosed within multiples of the half-light can be determined fairly accurately, whereas the total halo mass is only weakly constrained from measurements of $\langle \sigma^2_\mathrm{los}\rangle$ and $\Rh$ alone (see Fig.~\ref{fig:lambdaBias} for $\lambda \gg 1$).

We have tested the mass estimator on a suite of $N$-body mocks extracted from controlled re-simulations of the Aquarius A2 merger tree \citep[based on][]{EPLG17}, and recover the enclosed masses with an accuracy of $\lesssim 10$ per cent in systems with mass-loss fractions that differ by orders of magnitude. The re-simulations cover all subhaloes of the merger tree with masses at infall $\geq \unit[10^8]{M_{\odot}}$, and model each subhalo with the same number of $10^7$ $N$-body particles independent of its mass. This set-up allows us to follow the tidal evolution of subhaloes spanning many orders of magnitude of mass and size, limiting numerical issues like artificial disruption of poorly resolved low-mass substructures \citep[see][]{vdb2018}.
Motivated by the mounting observational evidence of DM cores in Milky Way dwarfs \citep[see e.g.][]{Walker2011,AmoriscoAgnelloEvans2013}, we run simulations assuming either cuspy or cored \citet{dehnen1993} DM density profiles for the subhaloes at infall.
Furthermore, we apply the minimum variance mass estimator to a catalogue Milky Way dwarf galaxies \citep{McConnachie2012}, showing a tight correlation between enclosed mass $M(<1.8\,\Rh)$ and luminosity $L_\star$. Using empirical functions (\emph{tidal tracks}) fitted to the evolution of stellar tracers embedded in the DM subhaloes of our re-simulations of the Aquarius A2 merger tree, we show that the correlation between enclosed mass and luminosity does not evolve significantly due to tidal stripping even when dwarfs lose more than one order of magnitude of their initial mass.
Our results suggest that the currently observed correlation is mainly driven by internal processes such as star-formation and feedback.
We furthermore find that the mean densities $\langle \varrho(<1.8\,\Rh)\rangle$ of ultrafaint galaxies are too high to be compatible with any of the simulated cored DM haloes, and show that tidal evolution further increases this discrepancy. This is a consequence of our choice of core size of the simulated DM haloes (a core size equal to the scale radius), and our findings suggest that the densities of ultra faint galaxies require core sizes that are much smaller than the scale radius of the DM halo. 

Constraints on the total halo mass from measurements of $\langle \sigma^2_\mathrm{los}\rangle$ and $\Rh$ alone are weak given our ignorance on how deeply embedded the stellar population is within the DM halo. For two-parameter DM density profiles - like the $\{\alpha,\beta,\gamma\}=\{1,5,0\}$ and $\{1,5,1\}$ profiles - the measurements of $\langle \sigma^2_\mathrm{los}\rangle$ and $\Rh$ however do constrain the DM halo structural parameters $\{\rmax,\vmax\}$ to follow a one-dimensional degeneracy curve which can be cast as a function of the segregation parameter $\Rh/\rmax$. Using our re-simulations of the Aquarius merger tree, we break this degeneracy by selecting those simulated DM subhaloes with $\{\rmax,\vmax\}$ values consistent with the degeneracy curve.{ This allows us to infer the total halo masses of Milky Way dwarf galaxies, assuming that the simulated subhaloes are representative of the population of Milky Way subhaloes. }
We have tested this method using mocks generated from re-simulations of the Aquarius merger tree and show that total halo masses are robustly recovered within a factor of the order of unity for both cuspy and cored DM haloes.
Our findings suggest that the classical Milky Way dSphs are embedded in haloes spanning a narrow range of masses, $8 < \declog(M/\mathrm{M_{\odot}}) < 10$, with no clear trend with either galaxy size or luminosity.
Surprisingly, we find that stellar mass and total halo mass of ultrafaint galaxies are \emph{anticorrelated}, i.e. the halo mass \emph{decreases} with increasing stellar mass.  
We caution that this anticorrelation may be caused by either observational inaccuracies (contamination by foreground stars, inflated velocity dispersions due to binary motion), or that the Aquarius A2 merger tree does not contain subhaloes representative of those of ultrafaint dwarfs.

With velocity measurements becoming available also tangentially to the line of sight \citep[e.g.][]{Massari2018}, note that it is straightforward to extend the virial mass estimator to systems with full kinematic information,
as the spherical virial theorem reads (cf. equations~\ref{eq:virial} -- \ref{eq:potTerm}):
\begin{equation}
 \langle \sigma_\mathrm{los}^2 \rangle + \langle\sigma^2_\alpha \rangle+ \langle\sigma^2_\delta \rangle = {4 \pi G}~\int_0^\infty r \nu_\star(r) M(<r) \mathrm{d}r  ~,
\end{equation}
with $\langle\sigma^2_\alpha \rangle$ and $\langle\sigma^2_\delta \rangle$ denoting the luminosity-averaged squared velocity dispersions of the two velocity components tangential to the line of sight.
Motivated by the new orbital constraints for Milky Way dSphs \citep{HelmiGaia2018} and ultrafaints \citep{Simon2018Gaia} available thanks to the Gaia satellite, in a future contribution, we will study how to make use of orbital motion and internal kinematics to further constrain the properties of DM haloes of Milky Way dSphs and their accretion histories.

\subsection*{Acknowledgements}
\urlstyle{rm}
This work used the ARCHER UK National Supercomputing Service (\url{http://www.archer.ac.uk}), and the authors would like to thank the administrators for their support. We thank the VIRGO Consortium for giving us access to the Aquarius trees. {The authors thank the anonymous referee for the detailed and valuable feedback.}
RE acknowledges support through the Scottish University Physics Alliance.

\footnotesize{
\bibliography{virial}
}

\appendix

\section{Tidal evolutionary tracks}
\label{appendix:tidal}
\begin{figure} 	 
  \centering
  \includegraphics[width=8.5cm]{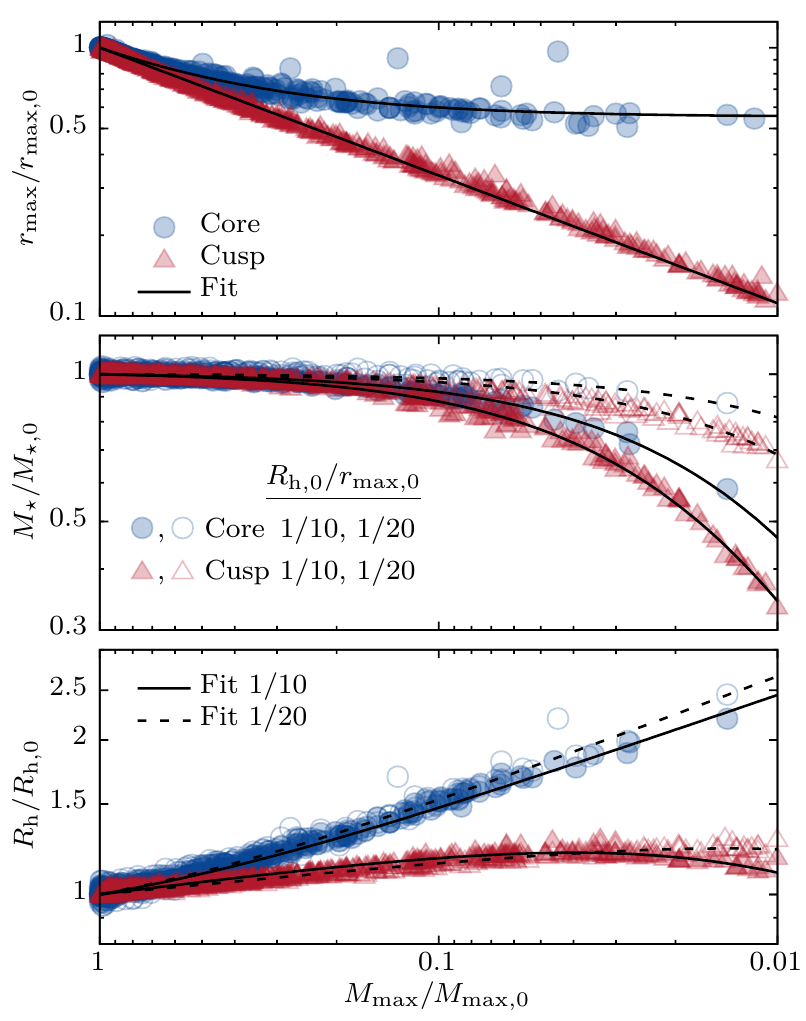} 
     \caption{Structural properties of dwarf galaxies are simple functions of the remnant bound mass only. The top panel shows $\rmax$ of the underlying DM halo, normalized by its initial value $r_\mathrm{max,0}$, as a function of the mass $M_\mathrm{max}/M_\mathrm{max,0}$ enclosed within $\rmax$. The evolution of $\rmax$ is shown separately for cuspy (red triangles) and cored (blue circles) DM profiles.
     The central and bottom panel show the evolution of the central surface brightness and half-light radii of stellar populations embedded inside the DM haloes with initial segregation of $R_\mathrm{h,0}/r_\mathrm{max,0} = 1/10$ (filled symbols), $1/20$ (open symbols).}
       \label{fig:tidaltracks}
\end{figure}

Structural parameters of dwarf galaxies that are being tidally stripped can be parametrized as simple analytical functions which depend only on the total fraction of mass lost, but not on the specific orbit of the dwarf, or the host potential. This has been first discussed by \citet{penarrubia2008} for the case of spherical stellar systems embedded in cuspy DM haloes, who named these analytical functions \emph{tidal evolutionary tracks}. Similarly, \citet{EPT15} discuss tidal tracks for spherical dwarf galaxies with cored DM haloes, whereas \citet{Sanders2018} study dwarf galaxies where both the stellar tracer and the DM halo are flattened. 
In the following, we derive such tracks for DM subhaloes and dwarf galaxies in the re-simulations of the Aquarius A2 merger tree discussed in section \ref{sec:methodcompenclosed}. 
In contrast to the tracks of \citet{penarrubia2008}, instead of fitting the evolution of single galaxies on different orbits at subsequent apocentres, we fit tracks to the entire population of simulated dwarf galaxies at $z=0$. 
We fit the DM halo radius of maximum circular velocity $\rmax$, the total stellar mass of the embedded stellar population $M_\star$ and the half-light radius $\Rh$, normalized by their initial values (i.e. at $z_\mathrm{infall}$), as functions of the remnant mass fraction $M_\mathrm{max}$ enclosed within $\rmax$. This parametrisation is motivated by the fact that the enclosed mass $M_\mathrm{max}$ can be measured directly from the $N$-body snapshots of the simulation, without having to assume a specific mass profile. For the fits, we adapt the empirical formula of \citet{penarrubia2008} introducing as third parameter the scale $x_s$: 
\begin{equation}
\label{equ:tidaltrackbase}
 g(x) =  \frac{(1+x_s)^\alpha~x^\beta}{(x+x_s)^\alpha} ~~~~~\mbox{with}~~~~ x = M_\mathrm{max} /  M_\mathrm{max,0} ~~,
\end{equation}
where $g(x) = \rmax(x) / r_\mathrm{max,0}$, $M_\star(x) / M_{\star,0}$ or $\Rh(x) / R_\mathrm{h,0}$, respectively, {subscript zero denoting values at infall}.
We introduce the scale $x_s$ to use the same parametrisation $x = M_\mathrm{max} /  M_\mathrm{max,0}$ for both the DM- and the stellar tracks: the choice of $x_s=1$ in the formula of \citet{penarrubia2008} was adapted for tidal tracks parametrized by mass lost within the initial half-light radius, but does not yield reasonable fits for structural parameters of the stellar population when expressing the tidal tracks as functions of $x = M_\mathrm{max} /  M_\mathrm{max,0}$.
We do not fit the scale parameter $x_s$, but where the functional shape of $g(x)$ requires it, we use as scale the initial fraction of mass enclosed within the initial half-light radius:  $x_s = ( 1 + a_0/R_\mathrm{h,0})^{\gamma-3}$, where by $a_0$ we denote the \citet{dehnen1993} scale radius of the halo at infall. This choice reduces the dependence of the fits on segregation, and motivated the choice of $x_s=1$ in the stellar tracks derived by \citet{penarrubia2008}. We fit tidal tracks separately for cuspy ($\gamma=1$) and cored ($\gamma=0$) dwarf galaxies which at $z_\mathrm{infall}$ had \citet{dehnen1993} density profiles, i.e. $\{\alpha,\beta,\gamma\}=\{1,4,\gamma\}$. Fits are done separately for two stellar populations with initial segregations of $R_\mathrm{h,0}/ r_\mathrm{max,0} = 1/20$ and $1/10$. 
To limit the impact of numerical artefacts caused by the spatial resolution of the particle mesh code, the highest resolving co-moving grid having a spacing of $r_\mathrm{max,0}/128$, we include only haloes in the fits with $\rmax/r_\mathrm{max,0} > 1/10$. We only consider bound $N$-body particles for the fits, and fit only haloes where both the DM and the stellar tracer are approximately in virial equilibrium, requiring that $|2K+W|/(2K-W) < 0.2$ for the DM enclosed within $\rmax$, and $|2K_\mathrm{los}+W_\mathrm{los}|/(2K_\mathrm{los}-W_\mathrm{los}) < 0.05$ for the stars. 
Halo properties are then averaged in logarithmically spaced bins spanning $0.01 \leq M_\mathrm{max} /  M_\mathrm{max,0} \leq 1$ to avoid giving different weight to various mass fractions based on the halo abundance at that mass fraction. 
The empirical fit parameters are listed in Table~\ref{tab:tidaltrack}.
Fig.~\ref{fig:tidaltracks} shows the fitted tidal tracks, as well as the parameters of single dwarf galaxies at $z=0$ used for the fit. Note that $\rmax$ of cored dwarf galaxies decreases less rapidly during tidal stripping than for their cuspy counterparts, and half-light radii of stellar populations embedded in cored DM haloes expand during tidal stripping.

\begin{table}
\centering
\caption{Empirical fit parameters for equation~\ref{equ:tidaltrackbase}. By $r_\mathrm{max/0} \equiv \rmax/r_\mathrm{max,0}$ we denote the radius of maximum circular velocity of the DM halo at $z=0$, normalized by its initial value. Similarly, $M_{\star/0}\equiv M_{\star}/M_{\star,0}$ and $R_\mathrm{h/0}\equiv R_\mathrm{h}/R_\mathrm{h,0}$ are the total stellar mass and half-light radius of stellar populations embedded inside the DM halo, normalized by their respective initial value.
We separately list parameters for two stellar populations with initial segregations of $R_\mathrm{h,0}/r_\mathrm{max,0} = 1/20$ and $1/10$.  }
\label{tab:tidaltrack}

\begin{tabular}{@{\hskip 0.1cm}c@{\hskip 0.3cm}c@{\hskip 0.4cm}c@{\hskip 0.3cm}c@{\hskip 0.3cm}c@{\hskip 0.4cm}c@{\hskip 0.3cm}c@{\hskip 0.3cm}c@{\hskip 0.1cm}}
\toprule
                                               &                                   & \multicolumn{3}{c@{\hskip 0.7cm}}{\textsc{Cusp}} & \multicolumn{3}{c@{\hskip 0.4cm}}{\textsc{Core}} \\                                          
                                               &${R_\mathrm{h,0}}/{r_\mathrm{max,0}}$  & $\alpha$    & $\beta$   &$\declog\, x_s$ & $\alpha$    & $\beta$ &$\declog\, x_s$   \\   \cmidrule(l{0cm}r{0.5cm}){2-2}  \cmidrule(r{0.5cm}){3-5} \cmidrule(r){6-8} 
$r_\mathrm{max/0}$                             & --          & 0.00      & 0.48    &0.00  &   -0.85     &   0.00     &0.00        \\ [0.3cm]
\multirow{2}{*}{$M_{\star/0}$}                 & 1/20        & 1.87      &1.87     &-2.64  &   2.83      &   2.83     &-3.12       \\
                                               & 1/10        & 1.80      &1.80     &-2.08  &   2.05      &   2.05     &-2.33        \\[0.3cm]
\multirow{2}{*}{$R_\mathrm{h/0}$}              & 1/20        & 0.47      &0.41     &-2.64  &  -0.25      &  -0.25     &0.00        \\
                                               & 1/10        & 0.50      &0.42     &-2.08  &  -0.23      &  -0.23     &0.00      \\ \bottomrule
\end{tabular}
\end{table}

\bsp	

\label{lastpage}
\end{document}